\documentclass[12pt,preprint]{aastex}

\usepackage{graphicx}

\begin{document}

\title{Astrometric Detection of Terrestrial Planets in the Habitable Zones of Nearby Stars with SIM PlanetQuest}
\author{Joseph Catanzarite, Michael Shao, Angelle Tanner,
Stephen Unwin, and Jeffrey Yu} \affil{Jet Propulsion Laboratory,
California Institute of Technology, 4800 Oak Grove Drive, Pasadena,
CA 91109-8099 } \email{jcat@s383.jpl.nasa.gov}

\begin{abstract}

SIM (Space Interferometry Mission) PlanetQuest is a space-borne
Michelson interferometer for precision stellar astrometry, with a
nine meter baseline, currently slated for launch in 2015. One of the
principal science goals is the astrometric detection and orbit
characterization of terrestrial planets in the habitable zones of
nearby stars. Differential astrometry of the target star against a
set of reference stars lying within a degree will allow measurement
of the target star's reflex motion with astrometric accuracy of
$\sim1$~$\mu as$ in a single measurement.

The purpose of the present paper is to quantitatively assess SIM's
capability for detection (as opposed to characterization by orbit
determination) of terrestrial planets in the habitable zones of
nearby solar-type stars. Note that the orbital periods of these
planets are generally shorter than the five-year SIM mission. We
formulate a ``joint periodogram'' as a tool for planet detection
from astrometric data. For adequately sampled orbits, i.e., five or
more observations per period, over a sampling timespan longer than
the orbit period, we find that the joint periodogram is more
sensitive than the $\chi^2$ test for the null hypothesis. In our
analysis of the problem, we use Monte Carlo simulations of orbit
detection, together with realistic observing scenarios, actual
target and reference star lists, realistic estimates of SIM
instrument performance and plausible distributions of planetary
system parameters.

Performance is quantified by three metrics: minimum detectable
planet mass, number and mass distribution of detected planets, and
completeness of detections in each mass range.

We compare SIM's performance on target lists optimized for the SIM
and Terrestrial Planet Finder Coronograph (TPF-C) missions. Finally,
we discuss the issue of confidence in detections and non-detections,
and show how information from SIM's planet survey can enable TPF to
increase its yield of terrestrial planets.

\end{abstract}


\section{Introduction}\label{sec:Introduction}
SIM will be the first instrument to use astrometry to detect and
characterize terrestrial planets. The ability to make these
detections depends on the performance of the instrument, the details
of the observing scenarios, and the analysis of the astrometric
data. This paper presents detailed simulations of the planet
detection process. It includes a realistic model of the instrument
performance, realistic observations of target and reference stars,
and plausible distributions of the astrophysical parameters defining
the ensembles of planetary systems. SIM is now at the end of NASA's
Phase B development. All significant technologies have been verified
in laboratory testbeds, and the project is on track to build the
flight instrument for launch as early as 2011. The instrument model
in this paper is based on results from these testbeds.

Section \ref{sec:ObservingScenario} presents a brief description of
SIM's narrow angle observing scenario. Section
\ref{sec:HabitableZone} contains a discussion of the habitable zone
and shows how the astrometric signature of a planet in the habitable
zone of a main-sequence star scales with stellar luminosity and
distance. In Section \ref{sec:TargetList}, we describe lists of SIM
target stars and their characteristics, along with several possible
survey modes which trade off number of stars versus number of
observations per star. Section \ref{sec:JointPeriodogram} provides a
description of the joint periodogram technique for detection of
periodicities in astrometric data. We describe the methodology of
our study in Section \ref{sec:Results1}. The main results are
presented in Sections \ref{sec:Results1} and \ref{sec:Results2}, in
which we quantitatively characterize SIM's sensitivity for detection
of terrestrial planets, the expected mass distribution and total
number of terrestrial planets SIM will discover, and the
completeness of detected planets as a function of planet mass. In
Section \ref{sec:Enrichment} we briefly consider how SIM's
discoveries can benefit the TPF mission.

\section{Narrow-angle observing scenario} \label{sec:ObservingScenario}

For narrow-angle astrometry, a target star should be surrounded by a
group of reference stars located within a radius of about a degree.
Differential delay measurements of the target and reference stars
will be used to simultaneously estimate the position of the target
star with respect to the reference frame and remove the linear field
dependence in the delay measurements. (Yu 2002, Milman 2002, JPL
internal memos). The least squares problem involves three
parameters, requiring a minimum of three reference stars. Ideally
the target star is at the photo-center of the reference stars. For
non-ideal reference star positions, there is a penalty on the
position accuracy achievable on measurement of the target star with
respect to the reference frame. It is best to have more than three
reference stars per target star, since the more reference stars, the
closer their photocenter is to the target star, and thus the lower
the error penalty will be (assuming the distribution of reference
star candidates is uniform on the sky near a target star). Selection
of planet-search targets and their reference stars is discussed in
Section \ref{sec:TargetList}. We find that most SIM planet-search
targets have eight or more available bright K-giant reference star
candidates within a kpc. According to our own simulations, we expect
that four to six reference star candidates per target star will
survive a ground-based radial velocity vetting program before SIM
launches, and that of these, three or more will be astrometrically
clean, i.e. will have astrometric reflex motion below SIM's
detection threshold.

\begin{figure}[p!]
\includegraphics[scale=.5,angle=0]{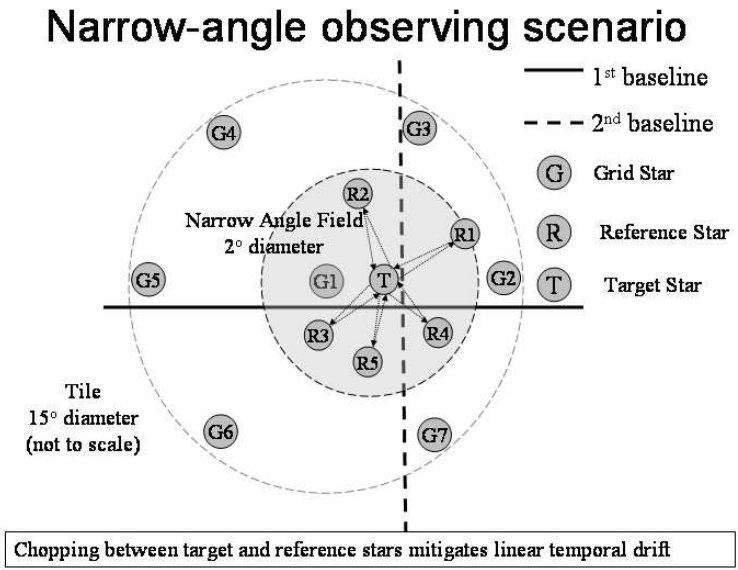}
\caption{Narrow-angle observing scenario for planet
surveys}\label{fig:30}
\end{figure}

For each visit to the target-reference group, the allocated
integration time is divided into a sequence of ``chopped''
measurements, alternating between target and reference stars. In
narrow-angle astrometry one is interested in the motion of the
target star with respect to the reference stars, rather than in the
absolute motion of the target star. Thus, a basic narrow-angle
measurement is always a difference between delay measurements of two
stars. Differences between successive target and reference star
measurements are, to first order, free of common-mode errors and of
linear temporal drift on the time scale of the chop. With optimally
selected integration time, chopping serves to mitigate systematic
time-dependent drifts (primarily due to changes in thermal
environment). A \emph{target-reference chop} is a 39-second delay
measurement on the target star, followed by a 39-second delay
measurement on the reference star. A \emph{chop cycle} is a complete
set of \emph{target-reference chops}. For five reference stars the
observing sequence for a \emph{chop cycle} is $ T\rightarrow R_{1}
\rightarrow T \rightarrow R_{2} \rightarrow T \rightarrow R_{3}
\rightarrow T \rightarrow R_{4} \rightarrow T \rightarrow R_{5}
\rightarrow$, where $T$ and $R_{i}$ refer to delay measurements of
the target star and $i^{th}$ reference star, respectively, and
$\rightarrow$ refers to a slew/settle/acquire sequence from either
target to reference star or reference to target star. The last
$\rightarrow$ is a slew/settle/acquire back to the target star.
Figure~\ref{fig:30} illustrates SIM's narrow-angle observing
scenario. The slew/settle/acquire time between stars in the
narrow-angle field is 15 seconds. A one-dimensional narrow-angle
observation sequence, or \emph{visit}, consists of two \emph{chop
cycles}, which comprises ten delay measurements on the target star,
a total of ten delay measurements on the reference stars, and twenty
slew/settle/acquisitions. At 39 seconds integration time per delay
measurement, and 15 seconds per slew/settle/acquire, this totals 780
seconds integration on the target and reference stars, plus 300
seconds slew/settle/acquisition time, for a total mission time of
1080 seconds, or 0.3 hours per visit to a target.

The astrometric precision obtained in this one-dimensional
narrow-angle observing sequence, for a bright ($V < 7$) target star,
or for the center of mass of the group of $V < 10$ reference stars,
is currently specified at 1.0 $\mu as$ \citep{AEB2005}. This is
subsequently referred to as SIM's \emph{single measurement
accuracy}. This performance has been demonstrated in the
Micro-Arcsecond Metrology (MAM) testbed at the Jet Propulsion
Laboratory, and has been accepted by the SIM External Independent
Review Team (EIRT). At 1.0 $\mu as$ single measurement accuracy, a
differential measurement with 780 seconds total integration time
divided among a $7^{th}$ magnitude target star and set of $10^{th}$
magnitude reference stars has astrometric accuracy of
$1.0\times\sqrt{2}~\mu as$. This includes photon noise, instrument
noise, and a multiplier that accounts for the geometric distribution
of the reference stars \citep{AEB2005}.

A two dimensional narrow-angle observation is a pair of
\emph{visits} (as described above) with the interferometer baseline
oriented along quasi-orthogonal directions on the plane of the sky.
For the first, and subsequent odd-numbered visits to the
target-reference group, the interferometer baseline is oriented
parallel to a reference direction on the plane tangent to the
spacecraft boresight direction. For the second, and subsequent
even-numbered visits to the target-reference group, the
interferometer baseline is oriented along a direction in the tangent
plane that is roughly orthogonal to the baseline orientation of the
first observation. In this way, the two-dimensional motion of the
target star on the plane of the sky is sampled. We assume even time
sampling for the series of observations along each of the two
baseline orientations, and that observation pairs are
quasi-simultaneous, although the latter assumption is not strictly
necessary.

In reality, scheduling constraints in the mission (including a solar
exclusion zone) preclude even sampling. Sampling of SIM
planet-search targets will be serendipitous, governed by their
availability during repeated ``orange-peel'' scans of the sky,
spiraling toward and away from the solar exclusion zone
\cite{B1997}. Yearly gaps in the sampling ranging from several weeks
to several months (for targets near the ecliptic) will occur during
times when the target is in the solar exclusion zone.

\cite{F2004, S2002} investigated planetary orbit detection with a
number of sampling schemes. These include geometric, power law, and
periodic with random gaussian perturbations, all with 24
two-dimensional observations. They found that all of these sampling
schemes performed well, as long as the minimum gap between
observations did not deviate too much from the average sampling
interval. The most promising observing schedules were `periodic with
perturbations' of up to 40\% of the average sampling interval
\cite{F2004}. Apart from annual sampling gaps due to the solar
exclusion-zone, we expect actual sampling to be quasi-even,
comparable to the `periodic with perturbation' schemes investigated
by \cite{F2004}. We have not investigated the impact of these solar
exclusion gaps on planet detectability. However, previous studies
\cite{F2004, S2002}, have shown that gaps much longer than the
average sampling interval can attenuate detection at periods
comparable to the survey length.

\section{Terrestrial planets and the habitable zone} \label{sec:HabitableZone}

Terrestrial planets are defined as those composed primarily of
silicate rock. In the solar system there are four (Mercury, Venus,
Earth, and Mars), Earth being the most massive. Recent simulations
of core accretion \citep{IL2004} indicate that rocky planets form
inwards of 3 AU from the parent star, and their masses can extend up
to about 10 or 20 $M_{\earth}$. The upper mass limit results from
the competition between core accretion and disk gas depletion. For
the purpose of this study, we adopt a terrestrial planet mass range
of one to ten $M_{\earth}$.

The habitable zone is the region around a star in which liquid
water, considered essential for life, can exist. Although life on
Earth exists in environments much more extreme than this definition
allows, feasible future missions such as the Terrestrial Planet
Finder will be limited to studying the macroscopic physical and
chemical properties of planets. From arguments based on Stefan's
Law, the Sun's habitable zone is between 0.7 and 1.5 AU
\citep{K1993}. For the purpose of this study, we put the center of
the Sun's habitable zone at 1 AU. To remain inside the habitable
zone, a planet with a semi-major axis of 1 AU should have
eccentricity less than 0.35.

We define the occurrence rate $\eta_{terrestrial}$ as the fraction
of solar-type stars with terrestrial planets orbiting in their
habitable zones. The mass distribution and occurrence rate of
extrasolar terrestrial planets orbiting solar-type stars are at
present unknown; only one candidate, at $\sim7.$5 $M_{\earth}$
\citep{R2005} has been discovered to date. NASA's Kepler mission
\citep{K2008}, scheduled to launch in 2009, will survey 100,000
solar-type stars (F, G and K dwarfs) over four years for transits of
planets with masses between 0.5 and 10 $M_{\earth}$. By the time of
the SIM launch, data from NASA's Kepler mission may have yielded
much information about the statistics of terrestrial planets
orbiting inward of $\sim1$ AU. In any case, SIM will itself provide
sufficient statistics to estimate the mass function and occurrence
rate of terrestrial planets in habitable zones.

At the present time, the best approach is to extrapolate from the
discoveries of radial velocity surveys. Masses of known extrasolar
planets are roughly consistent with a power-law distribution $dN/dM
\varpropto~M^{-1.1}$ \citep{TT2002, M2005C}, but they are generally
tens to hundreds of times more massive than terrestrial planets.
Nevertheless, the consistency of this power law with masses of solar
system planets is evidence (albeit weak) that it may also apply to
terrestrial planets \citep{TZ2003}. Integrating the power-laws for
mass and period distributions \cite{TT2002} between 1 and 10
$M_{\earth}$ and periods corresponding to orbit radii between 0.7 AU
and 1.5 AU, we obtain an estimate of $\eta_{terrestrial}$ = 0.013,
or 1.3\% for the occurrence rate of terrestrial planets in the
habitable zones of solar-type stars. In the remainder of this work,
we adopt the Tabachnik/Tremaine power law for terrestrial planet
masses, assuming that each star has one terrestrial planet at
mid-habitable zone, with mass drawn from the $1/M^{1.1}$
distribution. Results can then be scaled to any value of
$\eta_{terrestrial}$. In this work, we do not address the case of
multiple-planet systems.

The habitable zone of a star scales with luminosity as
\begin{equation}\label{eq:1}R_{H} = L_{\star}^{0.5},
\end{equation}where $R_{H}$ is the radius at mid-habitable zone in AU and $L_{\star}$ is stellar bolometric luminosity in
units of the solar bolometric luminosity $L_{\sun}$. The astrometric
signature of a planet in the habitable zone is
\begin{equation}
\label{eq:2} \alpha^{\prime\prime} =
\frac{M_p}{M_{\star}}\frac{R_{H}}{D},\end{equation} where
$\alpha^{\prime\prime}$ is the angular size of the semi-major axis
of the stellar reflex motion (if the orbit were seen face-on) in
arcseconds, $M_p$ and $M_{\star}$ are planetary and stellar masses
in solar units, respectively, and $D$ is distance to the star in pc.
Using Equation~\ref{eq:1} for $R_{H}$ in Equation~\ref{eq:2} gives
\begin{equation}
\label{eq:3} \alpha^{\prime\prime} =
\frac{M_p}{M_{\star}}\frac{L_{\star}^{0.5}}{D},\end{equation}

\noindent Most planet search targets are main sequence stars. A
convenient form of the mass-luminosity relation for main-sequence
stars with $M_{\star} > 0.2$ is \citep[p. 132]{AQ2000}
\begin{equation}
\label{eq:4} L_{\star} = M_{\star}^{3.8}.\end{equation}

\noindent Thus for main-sequence stars with $M_{\star} > 0.2$, the
radius of the habitable zone scales with stellar mass as
\begin{equation}
\label{eq:5} R_{H} = M_{\star}^{1.9}.\end{equation} More massive
stars have larger habitable zones. A consequence for TPF-C (but not
for SIM) is that for a fixed planet size, a larger habitable zone
lowers the contrast ratio between the planet's reflected starlight
and the star itself. Above a luminosity of $5.4~L_{\sun}$,
corresponding to a habitable zone of radius 2.3 AU, and a stellar
mass of 1.6 $M_{\sun}$, the contrast ratio of a 10 $M_{\Earth}$
terrestrial planet falls below TPF-C's contrast limit of $10^{-10}$.
From Equations~\ref{eq:3} and ~\ref{eq:4}, the astrometric signature
of a planet in the habitable zone of a main-sequence star with
$M_{\star} > 0.2$ scales with stellar mass, planet mass, and
distance as
\begin{equation}
\label{eq:6} \alpha^{\prime\prime} =
\frac{M_pM_{\star}^{0.9}}{D}\end{equation}

Evidently at fixed stellar distance, planets of a given mass in the
habitable zones of more massive stars have larger stellar reflex
motion signatures. These are the best targets for the SIM mission;
but some of them will be unsuitable for TPF-C because of low
contrast ratio. Targets for the TPF-C mission will be preferentially
selected for large habitable zone angular size, subject to the
contrast ratio constraint. In the next section, we discuss
hypothetical SIM target lists, and their characteristics. In our
simulations, we replaced Equation~\ref{eq:4} with a slightly more
accurate mass-luminosity relation \citep{GHM1988}, given by
\begin{equation}
\label{eq:7} \log_{10}L_{\star} = 4.20\sin(\log_{10}M_{\star} -
0.281) + 1.174, \end{equation} for $-1 < \log_{10}M_{\star} < 1.25$.

\section{Target lists for SIM planet surveys}\label{sec:TargetList}
Approximately 17\% of SIM's five-year mission
time is designated for planet-searching in narrow-angle mode. Within
this allocation, we consider three hypothetical survey modes, each
of which uses all of SIM's planet-finding time to observe targets
brighter than $7^{th}$ magnitude:

\noindent \textbf{Medium-Deep survey} -- 240 target stars with 52
two-dimensional observations per target.

\noindent \textbf{Deep Survey} -- 120 target stars with 104
two-dimensional observations per target.

\noindent \textbf{Ultra-Deep Survey} -- 60 target stars with 208
two-dimensional observations per target.

\noindent All observations in these surveys are made at SIM's
nominal single measurement accuracy of 1.0~$\mu as$.  For each
survey mode (except Medium-Deep, for which there are not enough
TPF-C targets) we draw the stars from one of two target lists. The
first is optimized for SIM, while the second is optimized for TPF-C.

Since virtually all likely SIM targets are known stars, it's
unnecessary to use synthetic stellar populations. Our hypothetical
SIM-optimized target list is derived from an initial list of 2350
stars taken from the Hipparcos catalog with distances less than 30
pc \cite{TT2003}. We removed all stars with $V > 7$, the limiting
magnitude for the integration time assumed in the narrow-angle
observing scenario. We removed those stars with stellar companions
within one arc-second of the primary star, to avoid contamination of
the primary star's fringe. We removed all stars with stellar
companions orbiting with semi-major axis within a factor of ten of
the radius of the mid-habitable zone of the primary star. This is a
conservative limit at which a companion will not have a significant
gravitational effect on a planet within the habitable zone
\cite{HW1999}. We screened out possible giant stars by the following
process. Stars catalogued as luminosity class III in SIMBAD were
removed if they had luminosity consistent with a giant star, but
kept if their luminosity was consistent with a dwarf. Stars
catalogued as dwarfs were removed if they had luminosity exceeding
that expected of a dwarf star. After these cuts, the 575 remaining
stars were sorted in descending order of the stellar astrometric
signature that would be induced by an Earth-mass planet at
mid-habitable zone. Finally, to optimize for detection of
terrestrial planets in the habitable zone, stars whose orbit periods
at mid-habitable zone were longer than five years or shorter than
0.2 years (the period corresponding to Nyquist sampling of 50
observations over five years) were removed from the list. After this
cut, the list contained 545 stars. The top 60, 120 and 240 stars on
this list are the targets for the Ultra-Deep, Deep and Medium-Deep
surveys, respectively.

For the hypothetical TPF-C targets, we used a list of 384 stars
\citep{B2005}. This list was derived from the Hipparcos database
~\cite{TT2003}. It comprises all single F, G, or K main sequence
stars brighter than $7^{th}$ magnitude, closer than 30 pc, and with
$B-V$ colors in the range 0.3 to 1.4. Two further constraints are
imposed: the mid-habitable zone must be outside of TPF-C's Inner
Working Angle of 62 mas, and the luminosity of the star must be less
than $5.4~L_{\sun}$, so that the contrast ratio exceeds $10^{-10}$
for a terrestrial planet of $10~M_{\earth}$. Of the 384 stars in the
list, about 120 survived application of these constraints, so for
the TPF-optimized list, there is no Medium-Deep survey. These stars
were then ranked in descending order of star-planet angular
separation for a planet at mid-Habitable zone. For the Ultra-Deep
and Deep TPF-C surveys we chose the best 60 and 120 stars,
respectively, from this final list. We note that these are not
necessarily the best targets for the TPF-Interferometer mission
(TPF-I). TFP-I has a smaller inner working angle, and could
potentially detect planets in the much smaller habitable zones of
M-stars.

\begin{figure}[p!]
\plottwo{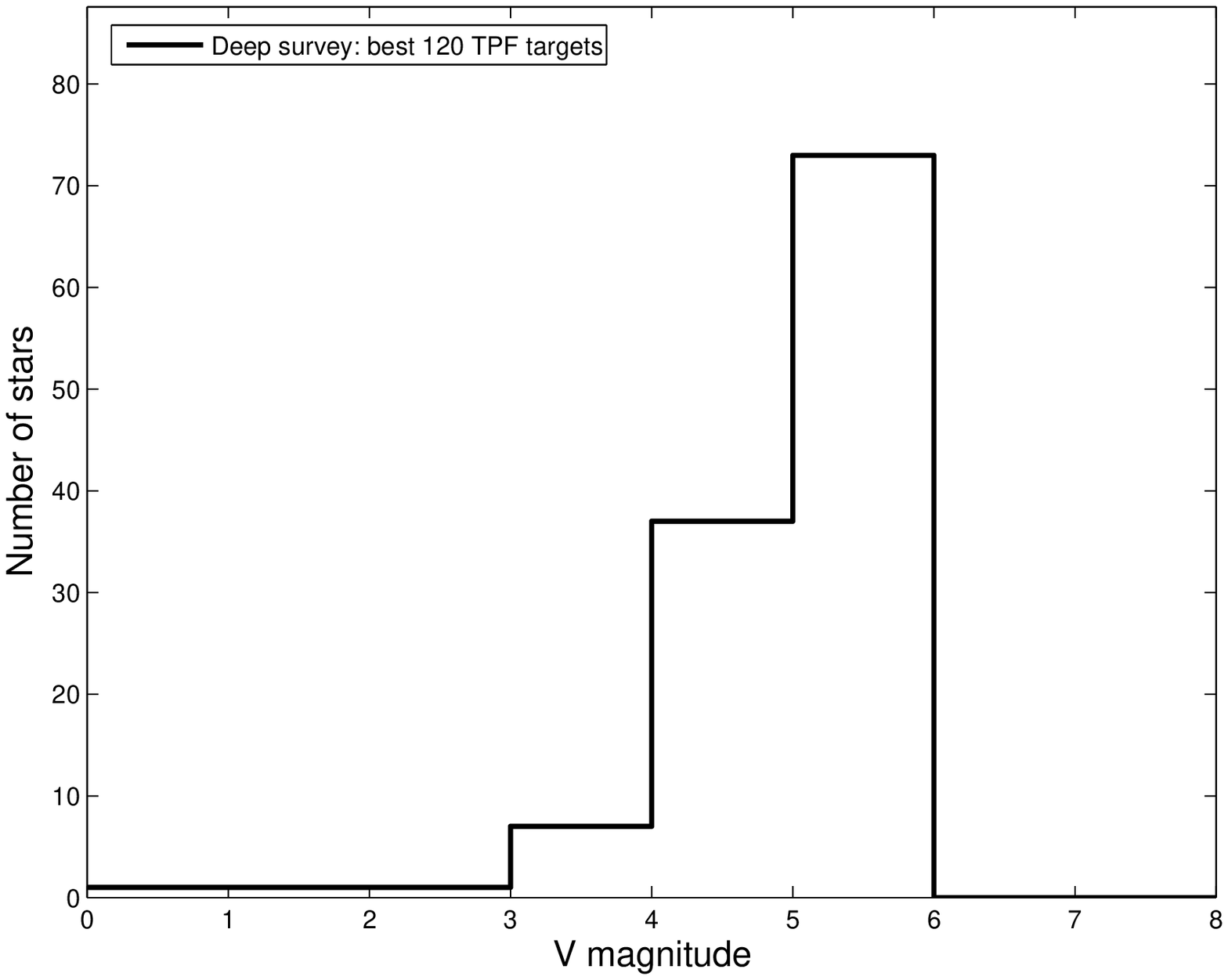}{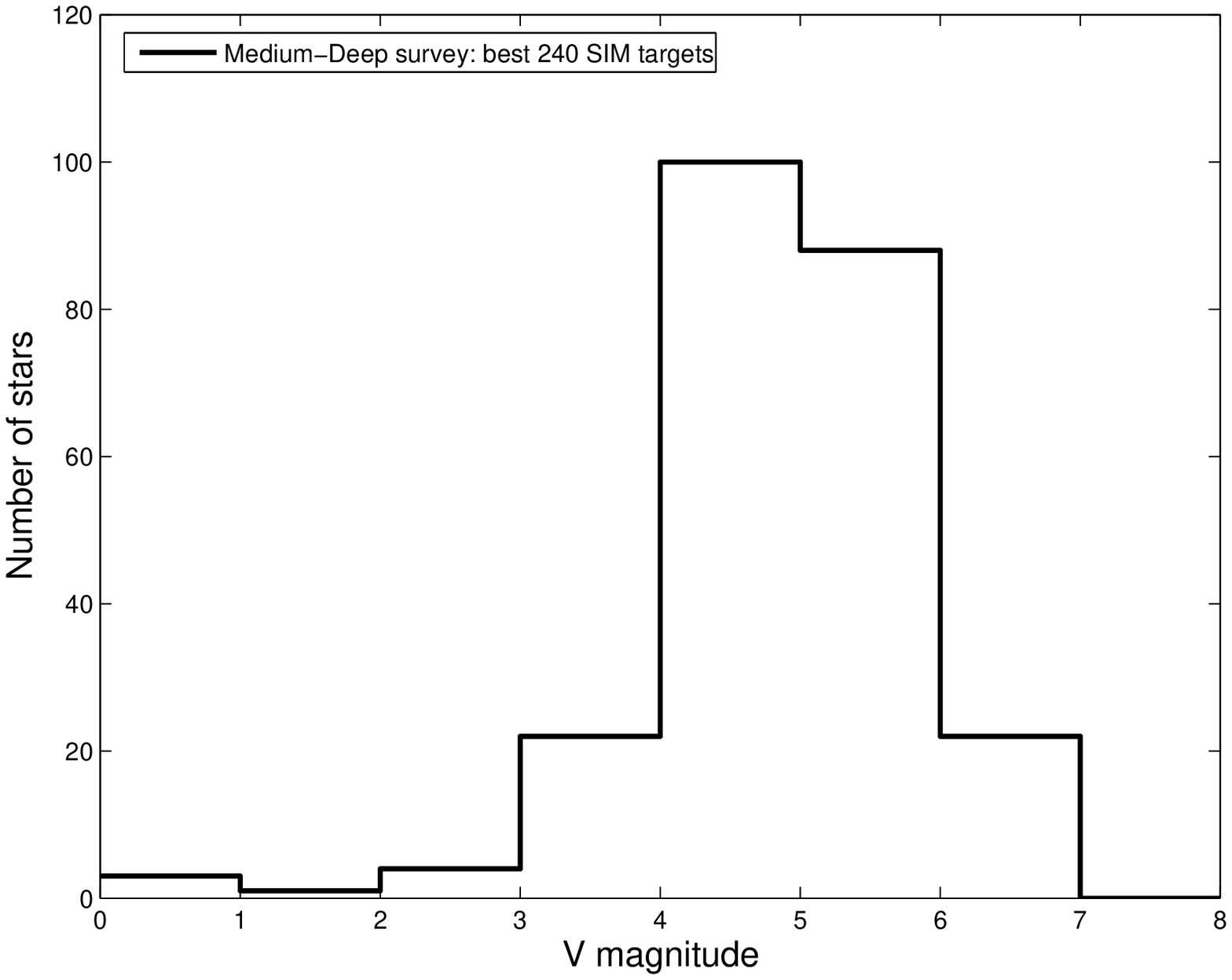} \caption{V magnitudes. Left:
Best 120 TPF target stars. Right: Best 240 SIM target
stars.}\label{fig:21}
\end{figure}

\begin{figure}[p!]
\plottwo{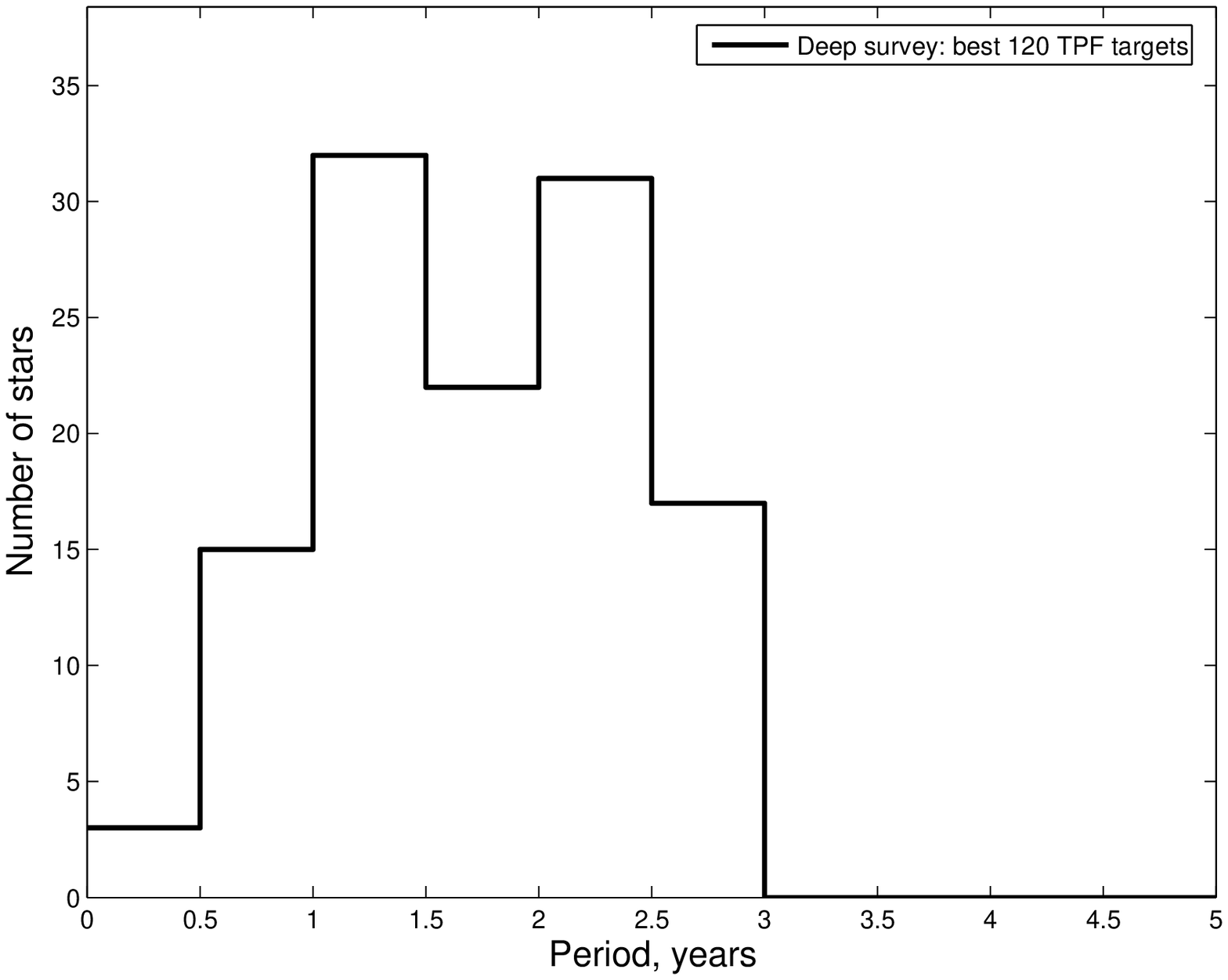}{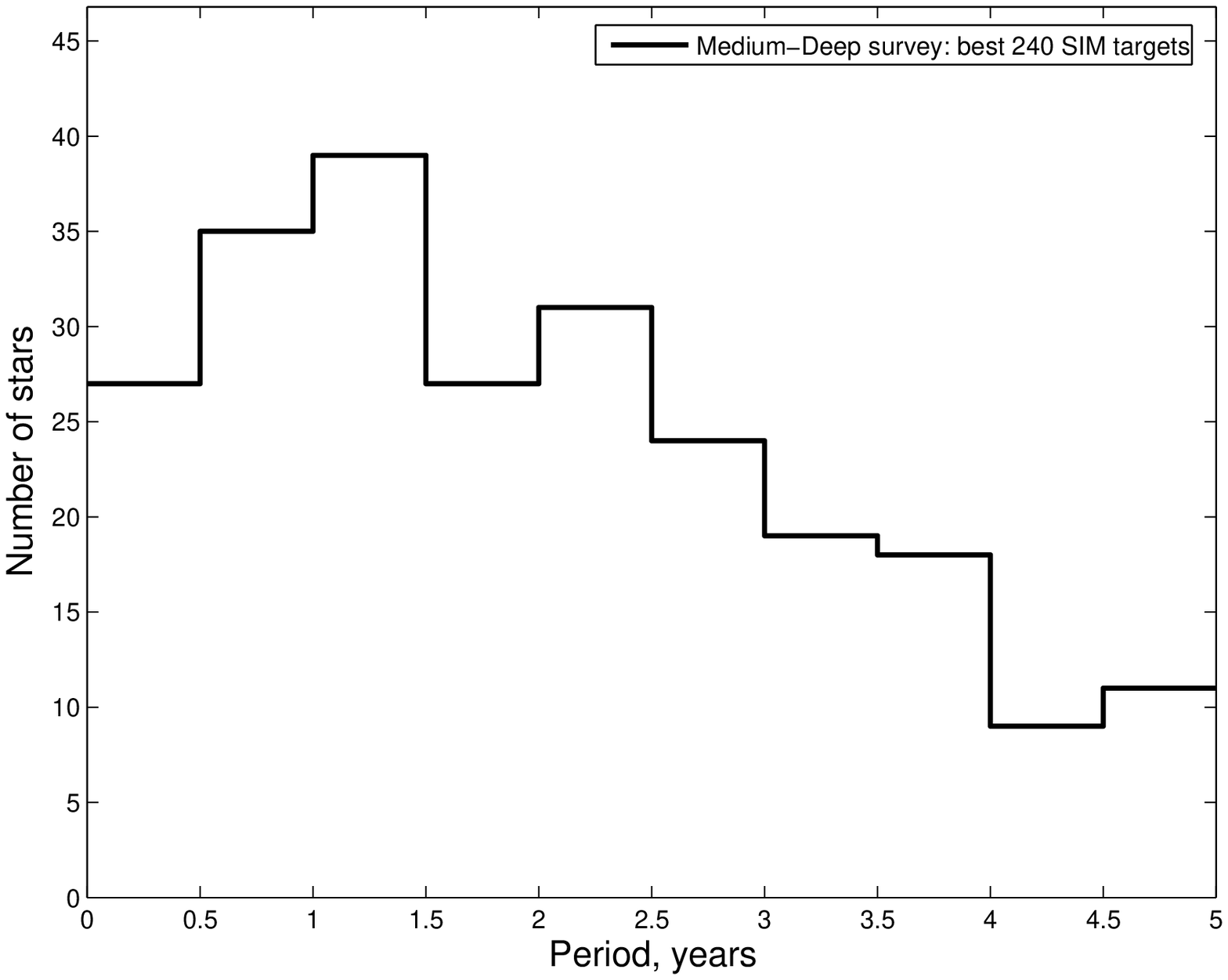} \caption{Orbital period at
mid-habitable zone. Left: Best 120 TPF target stars. Right: Best 240
SIM target stars. } \label{fig:22}
\end{figure}

Histograms of V magnitude and orbital period at mid-habitable zone
for the best 120 TPF-C targets and for the best 240 SIM targets are
shown in Figures \ref{fig:21} and \ref{fig:22}.  Note the wider
range of mid-habitable zone periods in the SIM target list, which
has been selected for large habitable zones.

Reference stars are K-giants selected from the Tycho 2 and 2MASS
catalogs \cite{G2006}. The following criteria were used:

\begin{enumerate}
\item They should be K Giant stars which are luminous and therefore distant, in order to minimize gravitational perturbations due to planetary companions. Reduced Proper
Motion, defined as $RPM = K_{S} + 5 log(\mu)$ serves as a proxy for
distance. We require $RPM < 1$.
\item They should have an infrared color range $0.5 < J - K_{S} < 1.0$
\item They should have an optical color range $1.0 < B - V < 1.5$.
\item They should lie within a 1.25 degree radius of the target star
\item They should be bright: $V < 10$ to minimize observing time
\item They should have favorable geometry: we choose four
reference stars in each quadrant around target if possible
\end{enumerate}

Our planet-search target and reference stars have not been screened
for photometric variability, which may be an indicator of starspot
activity. \cite{S2005, H2002} have modeled photocenter shifts due to
starspots; their results show that for typical planet search
targets, large spots causing a flux change of 1\% can induce
photocenter shifts of the same order as astrometric reflex motion
due to terrestrial planets in the habitable zone. \cite{S2005} notes
that since the effect correlates strongly with photometric
variability, photometric screening and/or monitoring of
planet-search targets should be considered. An average sunspot group
contains about ten sunspots, each with an area of about 0.04\% of
the Sun's disk.\cite{AQ2000}. The presence of simultaneous multiple
starspots tends to randomize the photocenter shift. For a solar-type
target star at 10 pc, we find, in agreement with \cite{S2005}, that
0.1\% flux variations due to a single starspot introduces a
photocenter shift with amplitude of $\sim0.3$~$\mu as$. While larger
starspots may be possible, we believe that since the photometric
shift due to starspots is color-dependent whereas reflex motion is
not, spectral information provided by SIM observations will serve to
break the degeneracy.

We consider also the problem of photocenter shifts in reference
stars. \cite{H2000} performed a survey of 187 F, G, and K giants for
photometric variability . They found that photometric variability
exhibits a strong color dependence; for giants meeting our selection
criterion of $1.0 < B - V < 1.5$, most exhibit maximum flux
variations of well under 1\%. They concluded that for giants cooler
than K2III, the observed photometric variation is most likely due to
radial stellar pulsations (which cause no photocenter shift) rather
than starspots. If, however, we assume that the observed photometric
variation is due entirely to starspots, a simple starspot model
shows that photocenter shift of K giants at 1 kpc is expected to be
well under 0.4 uas. Our conclusion is that starspots are not a major
concern for reference stars.

\section{The joint periodogram: a tool for astrometric detection of planets}\label{sec:JointPeriodogram}

The standard method of detecting periodicity in one-dimensional time
series data is the Lomb-Scargle periodogram \citep{S1982}. The
Lomb-Scargle periodogram and its variants are widely used in
detection of planets in radial velocity data  \citep{C2004, CM1999,
NA1998, W1995}. Discussion of the application of the periodogram to
detection of planets in astrometric data is found in \cite{BS1982},
and \cite{S2003} have investigated using the periodogram to detect
multiple planets in astrometric data.

The Lomb-Scargle periodogram is readily extended to detection of
periodicity in two-dimensional astrometric data time-series. We
define the joint periodogram as the sum of the Lomb-Scargle
periodogram power of the astrometric signal in the two independent
channels associated with the orthogonal baseline orientations of the
SIM interferometer. There is no requirement on the maximum time
interval between a measurement in one channel and the corresponding
measurement in the other channel. Detection of a planet is
registered when its joint power exceeds a detection threshold set
according to the desired false-alarm level.

Measurements in each channel are assumed uniformly sampled in time.
Comparison of many sampling schemes shows that apart from the
aliasing problem, even sampling is best for detection of orbits with
periods shorter than the length of the survey \cite{F2004}. In our
use of the periodogram, we have sidestepped aliasing effects by
counting any signal that exceeds the detection threshold as a
detection, regardless of whether the detected and actual periods
match. Thus our analysis should yield detection efficiencies
comparable to those found using other sampling schemes that are
optimized to reduce aliasing \cite{F2004, S2002}. Our use of even
sampling in the simulations is mainly for convenience in the
analysis.

The joint periodogram uses all the information in both channels of
the astrometric data for planet detection and period estimation. It
is ideally suited for detection of well-sampled circular orbits with
period shorter than the time baseline of the observations. For
elliptical orbits, detection efficiency is reduced for two reasons:
power leaks into the overtones of the orbit frequency, and for
certain orbit geometries, the full astrometric signature of the
stellar reflex motion is not observed. However, we find that these
effects are not significant for the relatively small eccentricities
($e < 0.35$) we consider. Our simulations show that the joint
periodogram detects Keplerian orbits with higher efficiency than
using separate periodograms on the two channels.

A necessary preliminary step in any detection scheme is to establish
the false-alarm probability (FAP) corresponding to a range of
detection thresholds. The false-alarm probability corresponding to a
given detection threshold (periodogram power) is the likelihood that
pure Gaussian noise would produce a periodogram peak whose power
exceeds the detection threshold. Lowering the detection threshold
raises the false-alarm probability. In our simulations, we
determined detection thresholds corresponding to various false-alarm
probabilities using 100,000 Monte Carlo realizations of the
no-planet case, where the simulated observations consist purely of
Gaussian measurement noise. For the purpose of discovering planets
in radial velocity data, a 1\% false-alarm probability is commonly
required \citep{M2005B}. Detection of a signal exceeding the
threshold corresponding to a 1\% false-alarm probability is said to
be at a 99\% significance level.

\begin{figure}[p!]
\epsscale{0.75}\plotone{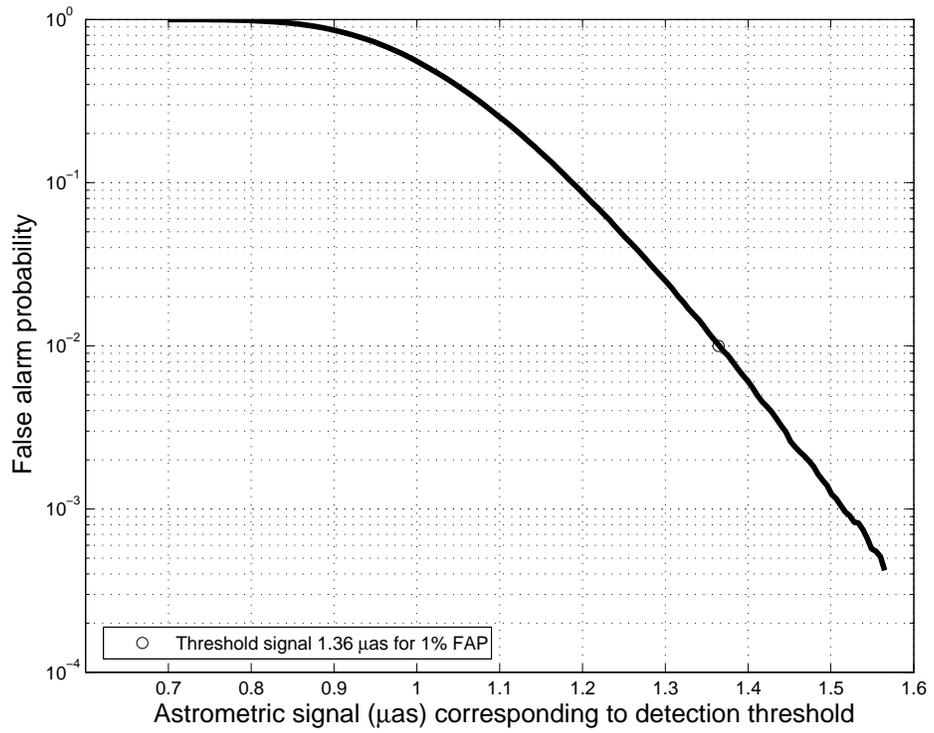} \caption{False-alarm probability
versus signal detection threshold for the joint periodogram: 100,000
trials, 50 two-dimensional observations per star, evenly sampled
over five years at single measurement precision of 1 $\mu as$.
}\label{fig:1}\end{figure}

For sufficiently small values, the false-alarm probability is
proportional to the number of independent frequencies scanned in the
periodogram (Horne \& Baliunas 1986; Press, et al. 1992). For N
evenly-spaced observations in the time series, the range of
detectable frequencies is spanned by the N/2 independent frequencies
\{ $\frac{1}{T}$, $\frac{2}{T}$ \ldots $\frac{N}{2T}$ \}, where T is
the time between the first and last measurement (nominally five
years for the SIM mission), and $\frac{N}{2T}$ is the Nyquist
frequency. But to adequately sample a peak, the periodogram must be
scanned at a finer frequency interval. Our Monte Carlo experiments
show, in agreement with Horne \& Baliunas (1986) and Press, et al.
(1992) that the effective number of independent frequencies is $\sim
N$, rather than $N/2$, because of this oversampling. Evidently,
limiting the range of frequencies over which the periodogram is
scanned increases the detection efficiency at a given confidence
level. For the results presented in this paper, we chose to sample
the periodogram down to periods as low as 0.2 years (semi-major axis
of 0.34 AU for a planet orbiting a solar-mass star) corresponding to
the Nyquist frequency for 50 samples evenly spaced in time over a
five-year mission. Figure~\ref{fig:1} is a plot of the FAP versus
detection threshold, from a Monte Carlo ensemble of 100,000
realizations of a series of 50 two-dimensional observations, evenly
spaced over five years of Gaussian noise, at $1~\mu as$
single-measurement error. The 1\% FAP threshold corresponds to
astrometric signal of 1.36 $\mu as$.

\section{Detection of terrestrial planets in the habitable zones of nearby stars with SIM} \label{sec:Results1}
In this section we determine SIM's sensitivity for detection of
terrestrial planets in the habitable zone, for hypothetical SIM and
TPF-C target lists. For each target list and survey strategy, we
answer the question: what is the distribution of minimum detectable
mass for terrestrial planets in the habitable zone?

We first generate Keplerian reflex motion orbits for a Monte Carlo
sample of solar-mass stars at 10 pc, each with a single terrestrial
planet (i.e. a planet with mass in the range 1 to 10 $M_{\Earth}$ ).
We determine the detection efficiency as a function of planet mass.
Then we extend this result, deriving a universal detection
efficiency curve for a star of arbitrary mass, luminosity and
distance, and observed with arbitrary number of two-dimensional
measurements and differential measurement error. From the universal
detection efficiency curve, we develop a semi-analytical formula for
minimum detectable mass of terrestrial planets in the habitable
zone. For each survey strategy (and its hypothetical target list),
we determine SIM's sensitivity for detection of terrestrial planets
in the habitable zone, for each target star.

These results provide a description of the sensitivity of the SIM
instrument for detecting terrestrial mass planets orbiting in the
habitable zones of nearby stars; they are independent of assumptions
concerning the occurrence rate and mass distribution of terrestrial
planets. Detailed results and discussion are presented in
\S\ref{sec:ResultsA}.

\subsection{Monte Carlo sample}
The starting point is a Monte Carlo simulation of astrometric
detection of planets in the habitable zone of a solar-type star at a
distance of 10 pc, in order to determine detection efficiency as a
function of planet mass. Detection efficiency at a given planet mass
is defined as the likelihood that a planet of that mass will be
detected in the presence of measurement noise. For each planet mass
in the range 0 to 10 $M_{\earth}$, at intervals of 0.5$M_{\earth}$,
we generate 10,000 realizations of Keplerian orbits with a one-year
period, around a solar-mass star at 10 pc. Eccentricity is uniformly
distributed between zero and a maximum of 0.35, consistent with
orbits lying entirely within the habitable zone. We randomly draw
other orbit parameters (inclination, longitude of ascending node,
longitude and time of periastron) from their allowed domains. For
each realization of an orbit, we generate a time series of 50
two-dimensional astrometric observations of the star's position,
evenly spaced in time over a five-year time period. Each observation
is perturbed with $1~\mu as$ Gaussian measurement error. In these
initial simulations, we did not account for the effects of parallax
and proper motion.

We employ the joint periodogram to detect periodic stellar reflex
motion indicating the presence of a planet. Detection efficiency at
a given mass and FAP is defined as the fraction of the ensemble of
Keplerian orbit realizations for which the joint periodogram power
exceeds the detection threshold associated with this FAP.

\begin{figure}[p!]
\epsscale{0.75}\plotone{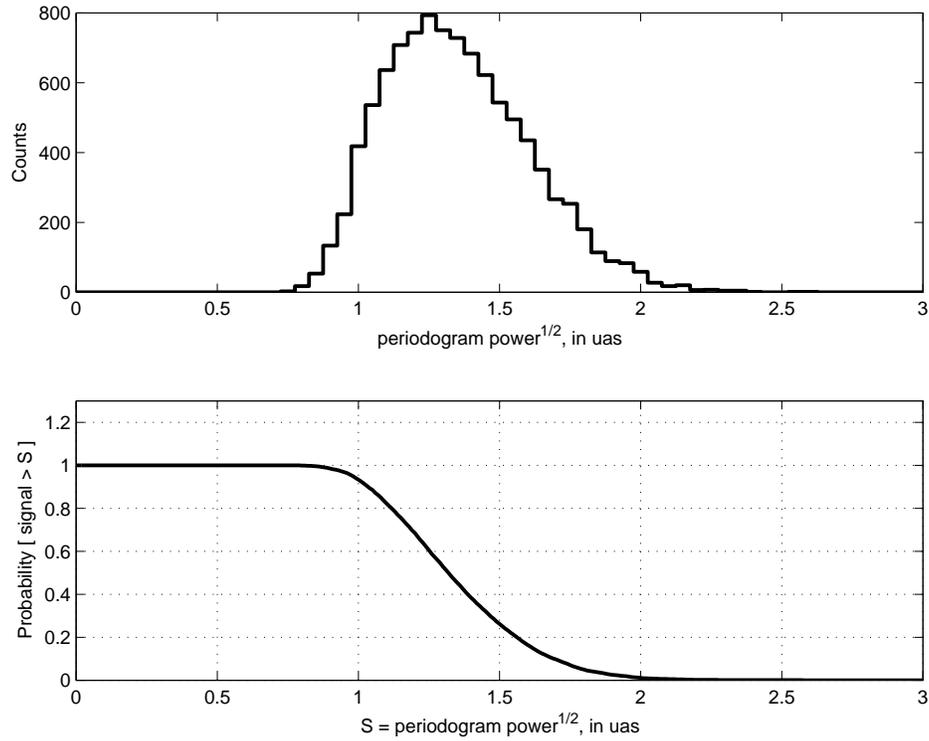} \caption{Upper panel: Periodogram
power distribution for an ensemble of 10,000 realizations of a 3.5
$M_{\Earth}$ planet in a one-year orbit around a solar-mass star at
a distance of 10 pc. Five-year mission, 50 evenly sampled
two-dimensional observations at 1 $\mu as$ single measurement
precision. Lower panel: Detection efficiency versus detection
threshold (normalized integral of the power distribution)}
\label{fig:2}\end{figure}

At moderate signal, a detection threshold corresponding to the peak
of the periodogram power distribution corresponds roughly to 50\%
detection efficiency -- noise is equally likely to add to the
signal, raising it above the threshold, or subtract from the signal,
pushing it below the threshold (since the distribution of
periodogram power is nearly symmetric). Increasing or decreasing the
detection threshold by the same increment results in equal and
opposite changes in the number of detected planets (see Figure~
\ref{fig:2}).

Figure~\ref{fig:3} shows SIM's detection efficiency (for detection
thresholds corresponding to false-alarm probabilities of 1\% and
5\%) as a function of planet mass for the fiducial case of
terrestrial planets in one-year orbits around a solar-type star at a
distance of 10 pc. Note that as required, in the limit as the signal
approaches zero, the detection efficiency approaches the false-alarm
probability.

\subsection{SIM planet detection sensitivity}
\label{sec:ResultsA} If we assume that detection efficiency is
independent of orbit period, then it should depend only on the
signal-to-noise ratio (SNR). This is expected to hold when there are
sufficiently many observations, the orbit period is shorter than the
observation time baseline, and the data do not include the effects
of parallax and proper motion. We define the SNR of the astrometric
data as
\begin{equation}\label{eq:8}
SNR = \frac{\alpha\sqrt{N_{obs}}}{\sqrt{2}\sigma},\end{equation}
where $\alpha$ is the angular size of the semimajor axis of the
orbit in $\mu as$, if it were viewed in a face-on orientation,
$N_{obs}$ is the number of two-dimensional observations of the star,
and $\sigma$ is the single-measurement accuracy in $\mu as$. A
factor of $\sqrt{2}$ occurs in the denominator because narrow-angle
measurements are differential (\S\ref{sec:ObservingScenario}). The
SIM technology testbeds have demonstrated that SNR improves with
$N_{obs}$ according to Equation (\ref{eq:8}) up to $N_{obs}
\geqslant 100$. Substituting for $\alpha$ using
$\alpha^{\prime\prime}$ from Equation~\ref{eq:3} (remembering to
convert from arcseconds to micro-arcseconds), converting planet mass
from solar-mass to earth-mass units, and using the mass-luminosity
relation of Equation~\ref{eq:4}, we have:

\begin{equation}\label{eq:9}
SNR = 3.004 \cdot M_p \frac{L_{\star}^{0.24}}{D}
\frac{\sqrt{N_{obs}}}{\sqrt{2}\sigma},\end{equation}

\noindent where $M_p$ is planet mass in units of $M_{\earth}$, and D
is the distance to the target star in pc.

\begin{figure}[p!]
\plotone{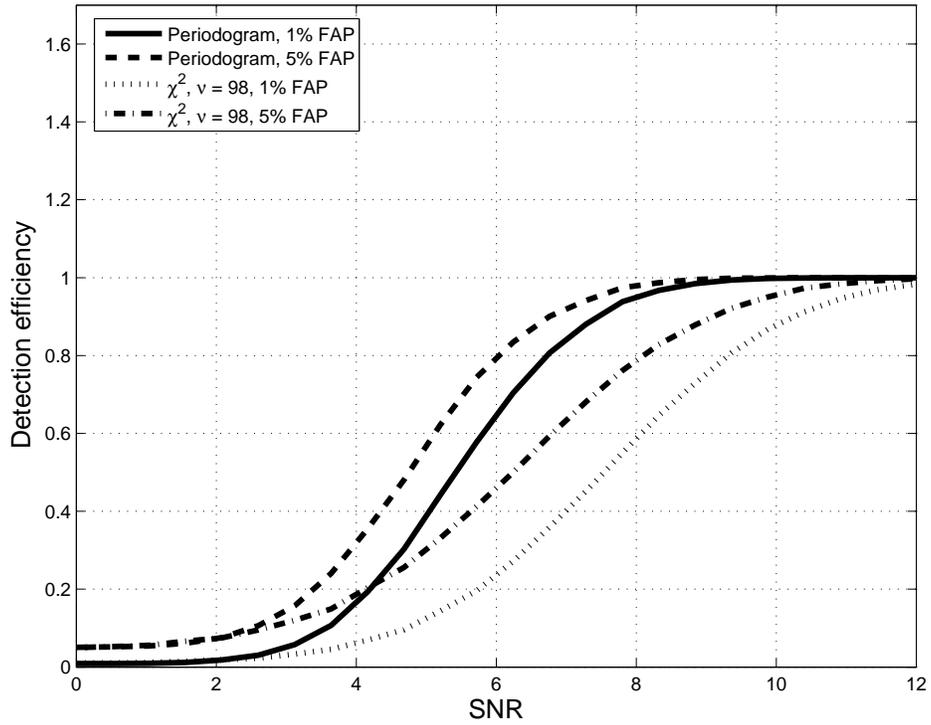} \caption{Universal SIM detection efficiency curves
for periodogram versus $\chi^{2}$ detection. Based on Monte Carlo
simulations with 50 2D observations of orbits with eccentricities in
the range $0 < e < 0.35$. Not corrected for the effects of proper
motion and parallax.} \label{fig:4}
\end{figure}

Universal detection efficiency curves comparing detection via the
joint periodogram and the $\chi^2$ test are shown in
Figure~\ref{fig:4}. For the $\chi^2$ test, detection is registered
when the $\chi^2$ test warrants a rejection of the null (no-planet)
hypothesis. The curves are derived from a Monte Carlo ensemble of
solar-mass stars at 10 pc, each with a planet in a one-year orbit,
with eccentricities in the range $0 < e < 0.35$, observed 50 times
along each of two orthogonal baseline directions with $1~\mu as$
single measurement error. We have confirmed via additional Monte
Carlo simulations for cases of 24, 100, and 200 two-dimensional
observations over five years, that detection efficiency versus SNR
curves are indeed identical.

We find that the joint periodogram is more sensitive than the
$\chi^2$ test of the null hypothesis for detection of planets in
Keplerian orbits (see Figure~\ref{fig:4}). Our simulations indicate
that this result is generally valid for well-sampled data, when the
orbit period is shorter than the survey duration. A reason may be
that since Gaussian noise has a flat frequency spectrum, the
$\chi^2$ test is sensitive to noise at all frequencies. By contrast,
periodogram power includes only noise at the natural frequencies
near the detected peak.

SIM planet detection sensitivity for a target can conveniently be
expressed in terms of minimum detectable planet mass. For the
purpose of discovering planets in radial velocity data, a 1\%
false-alarm probability (FAP) is commonly required \cite{M2005C}.
For each target star, we define a minimum detectable mass to be the
minimum mass of a planet orbiting in the habitable zone, detectable
at 1\% false-alarm probability with 50\% detection efficiency.

As an example of how detection efficiency and FAP are related,
consider Figure~\ref{fig:4}. At SNR of $\sim 5.4$, the detection
efficiency is 50\% at 1\% FAP, for planets with eccentricity
uniformly distributed in the range $0 < e < 0.35$. According to
Equation~\ref{eq:9}, SNR of $\sim 5.4$ corresponds to a minimum
detectable mass of $3.6M_{\earth}$, for 50 two-dimensional
observations of a solar-mass star at 10 pc, with single-measurement
precision of $1~\mu as$. Figure~\ref{fig:3} shows this graphically:
50\% detection efficiency occurs for a planet mass of
$3.6~M_{\earth}$. More generally, Equation~\ref{eq:9} evaluated at
SNR = 5.4 relates the minimum detectable mass to the number of
observations, single-measurement precision, distance and luminosity
of the star. Comparison with the case of 50 observations at $1~\mu
as$ single-measurement precision for a solar-type star at 10 pc
yields a useful semi-analytical scaling law for the minimum
detectable mass $M_{p,min}$ of a terrestrial planet in the habitable
zone of a main-sequence star:
\begin{equation}\label{eq:10}
    M_{p,min} =  3.6 M_{\Earth}\frac{\sigma}{1~\mu as} \sqrt{ \frac{50}{N_{obs}} } \frac{D}{10~pc} \frac {1}
    {L_{\star}^{0.24}},
\end{equation}

\noindent where $N_{obs}$ is the number of 2-dimensional
observations, $\sigma$ is the single-measurement accuracy in $\mu
as$, $D$ is the distance to the star in pc, and $L_{\star}$ is the
bolometric stellar luminosity in solar units.

\subsection{Correction for effects of parallax and proper motion}
\begin{figure}[p!]
\plotone{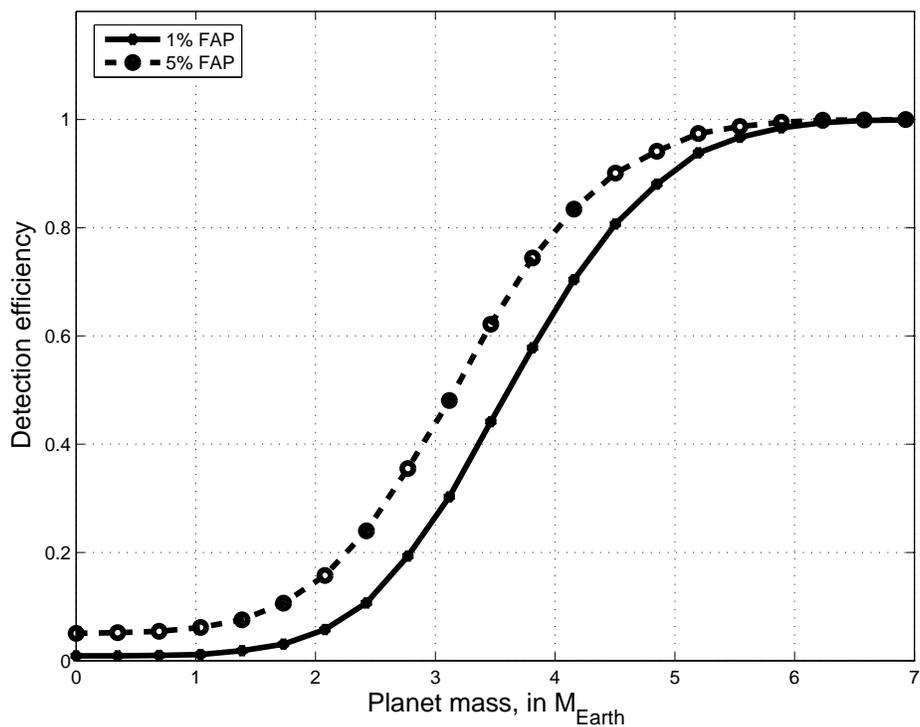} \caption{Detection efficiency versus planet mass
for a terrestrial planet in a one-year orbit around a solar-mass
star at a distance of 10 pc. Five-year mission, 50 two-dimensional
observations per star at single measurement precision of 1 $\mu as$;
orbital eccentricities in the range $0 < e < 0.35$. Each data point
represents an ensemble of 10,000 Monte Carlo orbits. Not corrected
for the effects of proper motion and parallax.} \label{fig:3}
\end{figure}

\begin{figure}[p!]
\plotone{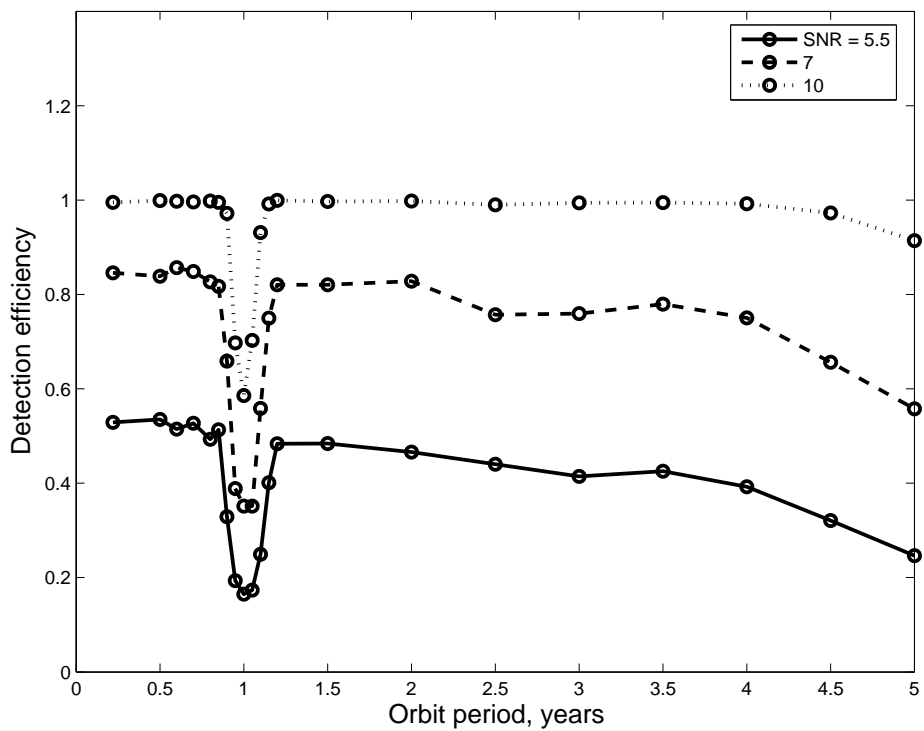} \caption{Sensitivity of periodogram detection to
period, for several values of signal-to-noise ratio (SNR). Each data
point represents an ensemble of 1000 Monte Carlo orbits. The loss in
sensitivity near orbit periods of one year is due to confusion of
stellar reflex motion with parallax; the steep decline at longer
orbit periods is due to confusion of stellar reflex motion with
proper motion. Detection threshold corresponds to 1\% false-alarm
probability.} \label{fig:5}
\end{figure}

In the foregoing development, we implicitly assumed that detection
efficiency is independent of orbit period. But when the effects of
parallax and proper motion are accounted for, detection probability
is attenuated for orbit periods approaching one year (due to
confusion of the orbit trajectory with parallax), and also for orbit
periods approaching the five-year observation window (due to
confusion of the orbit trajectory with proper motion). For
discussions and studies of these effects, see \cite{BS1982, L2000,
S2002, F2004}. We need to revise our approach in order to account
for sensitivity to orbit period. To characterize the effect, we
generated sets of Monte Carlo ensembles for the same range of planet
masses as reflected in Figure~\ref{fig:3}. Instead of putting every
star at 10 pc and every planet at orbit period of 1 year as before,
we generated ensembles of planets, with each ensemble having a fixed
orbit period chosen from a range of values between 0.2 and 5 years;
for each period we adjusted the distance to keep the astrometric
signal the same as if the planet were in a 1 AU orbit around a solar
mass star 10 pc away. We fitted the observations of each star to a
model for parallax (at ecliptic latitude of $30^{\circ}$) and proper
motion before running the periodogram search. Examples of correction
to detection efficiency as a function of period, for several values
of SNR are shown in Figure~\ref{fig:5}. The marked reduction in
detection efficiency for periods in the range 0.8 to 1.2 years is
due to confusion of stellar reflex motion with parallax. Confusion
of reflex motion with proper motion causes falloff in detection
efficiency with increasing period. To fully account for sensitivity
to period, we used the results of our Monte Carlo simulations to
construct a lookup table for detection efficiency as a function of
SNR and orbit period. For any target star, given the number of
observations and measurement noise, and the orbit period at its
mid-habitable zone, a detection efficiency curve (similar to
Figure~\ref{fig:3}, but corrected for the effects of parallax and
proper motion) is obtained by interpolation in this table.

\begin{figure}[p!]
\plotone{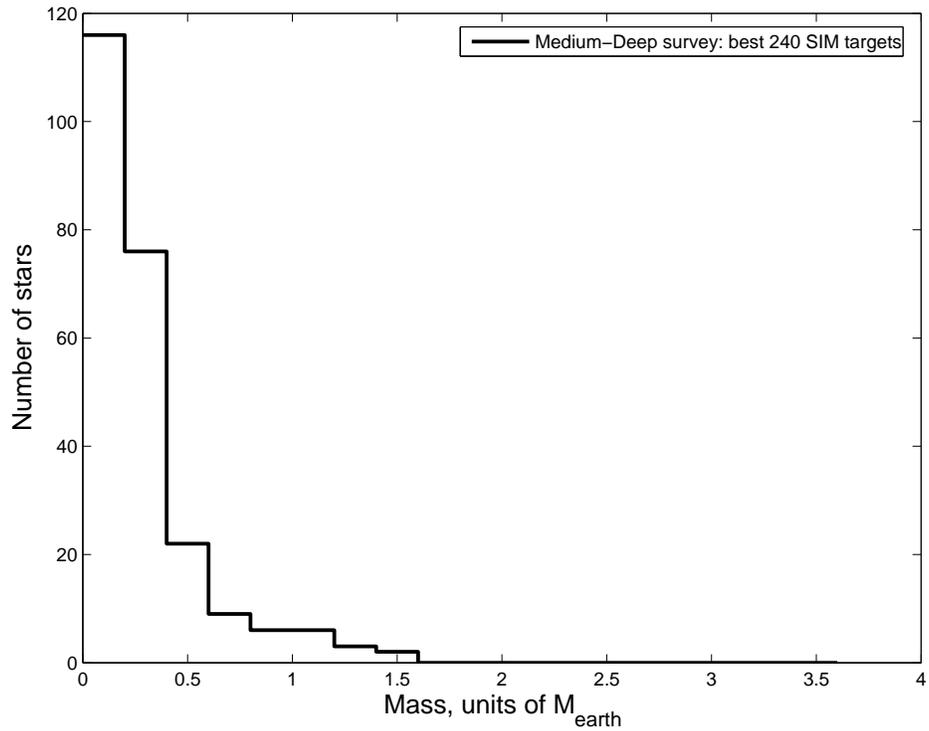} \caption{Correction to minimum detectable mass due
to parallax and proper motion. Medium-Deep planet survey -- best 240
stars for SIM, 52 two-dimensional measurements per star. Single
measurement precision is $1.0~\mu as$; minimum detectable mass is
for 50\% detection efficiency at detection threshold corresponding
to 1\% false-alarm probability.}\label{fig:40}
\end{figure}

Figure~\ref{fig:40} shows a histogram of the corrections to minimum
detectable mass for the Medium-deep survey; note that although for
some stars the correction can be quite large, the corrections for
most stars are $< 0.25M_{\earth}$.

From the detection efficiency curve for any target star (now
corrected for the effects of parallax and proper motion) we can
determine the minimum detectable mass of a terrestrial planet in the
habitable zone of that star. In all plots and tables to follow,
minimum detectable mass has been corrected for the effects of
parallax and proper motion.

\subsection{Results for minimum detectable planet mass}
By the time SIM launches in 2015, the ubiquity of terrestrial
planets orbiting solar-type stars may already be known from
discoveries of the Kepler mission. For a fixed amount of SIM mission
time, the optimum number of stars to survey for planets with SIM
depends on the abundance of Earth-mass planets in the local Galactic
environment. For example, if terrestrial planets turn out to be
relatively rare, a reasonable strategy for SIM is to survey a larger
number of stars with correspondingly fewer observations per star. To
explore this trade, we consider the Medium-Deep, Deep, and
Ultra-Deep survey modes, described in \S\ref{sec:TargetList}, which
use the same amount of SIM mission time to observe different numbers
of targets.

\begin{figure}[p!]
\epsscale{1}\plotone{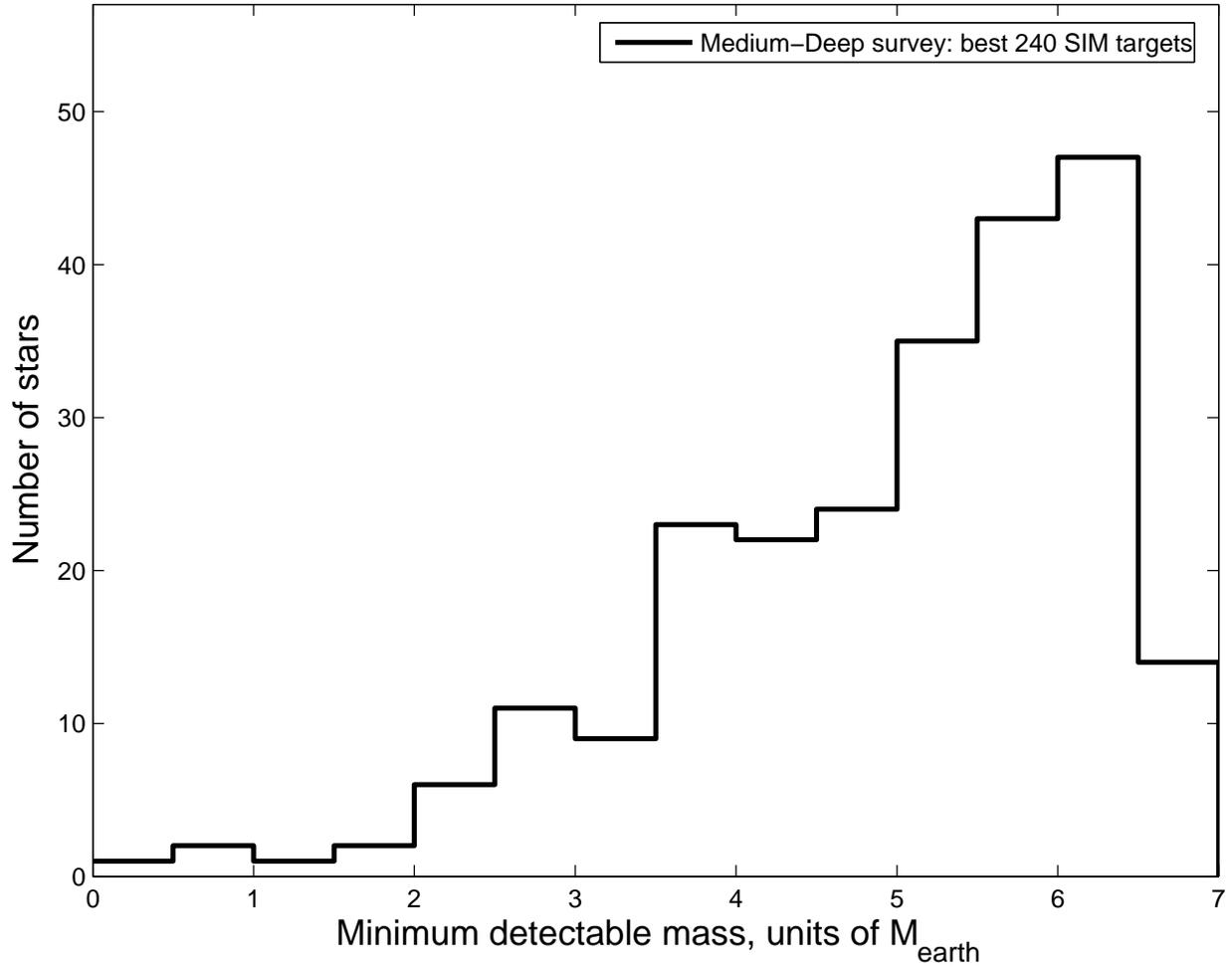} \caption{Distribution of minimum
detectable masses. Medium-Deep planet survey -- best 240 stars for
SIM, 52 two-dimensional measurements per star. Single measurement
precision is $1.0~\mu as$; minimum detectable mass is for 50\%
detection efficiency at detection threshold corresponding to 1\%
false-alarm probability.} \label{fig:6}
\end{figure}

\begin{figure}[p!]
\epsscale{1}\plottwo{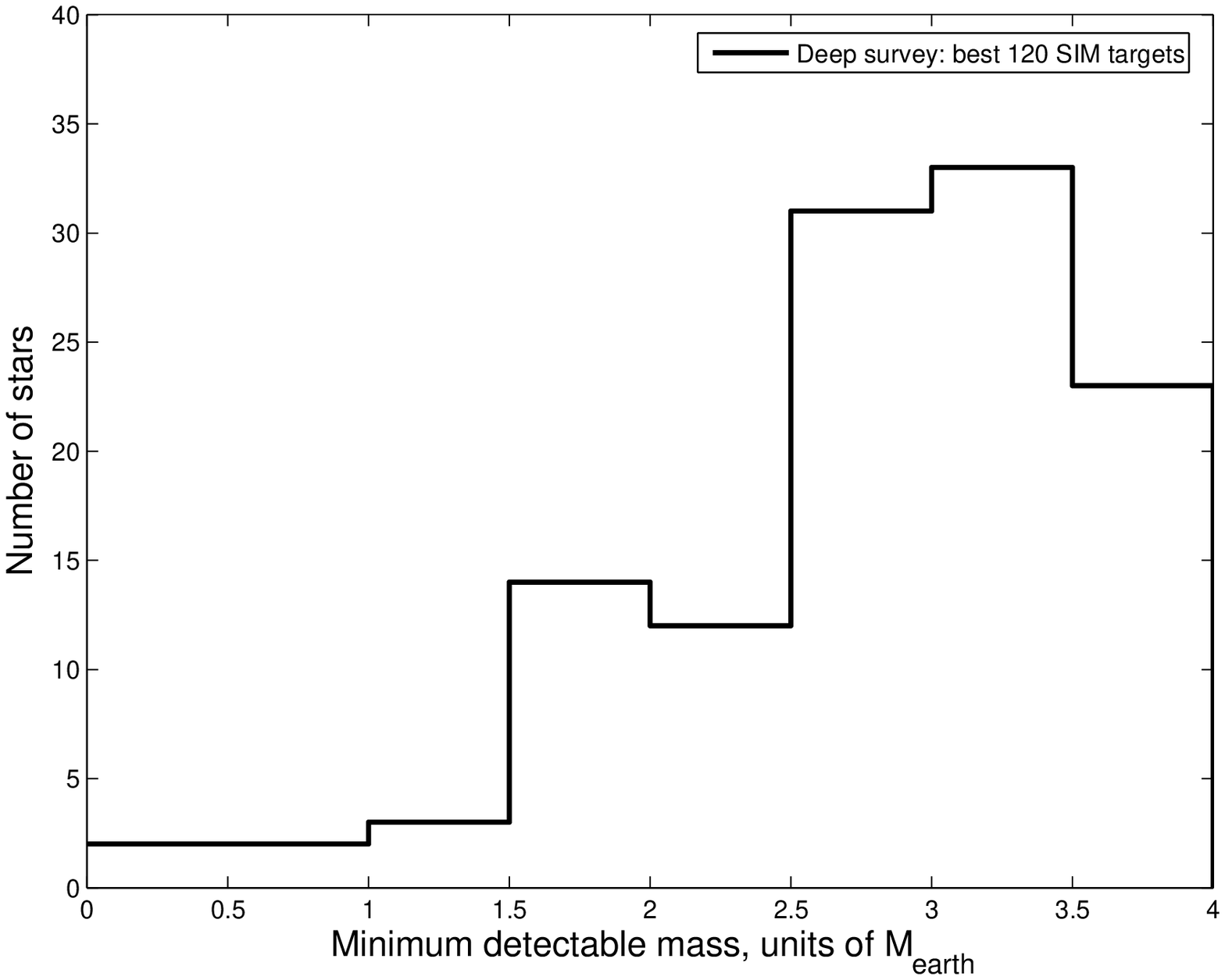}{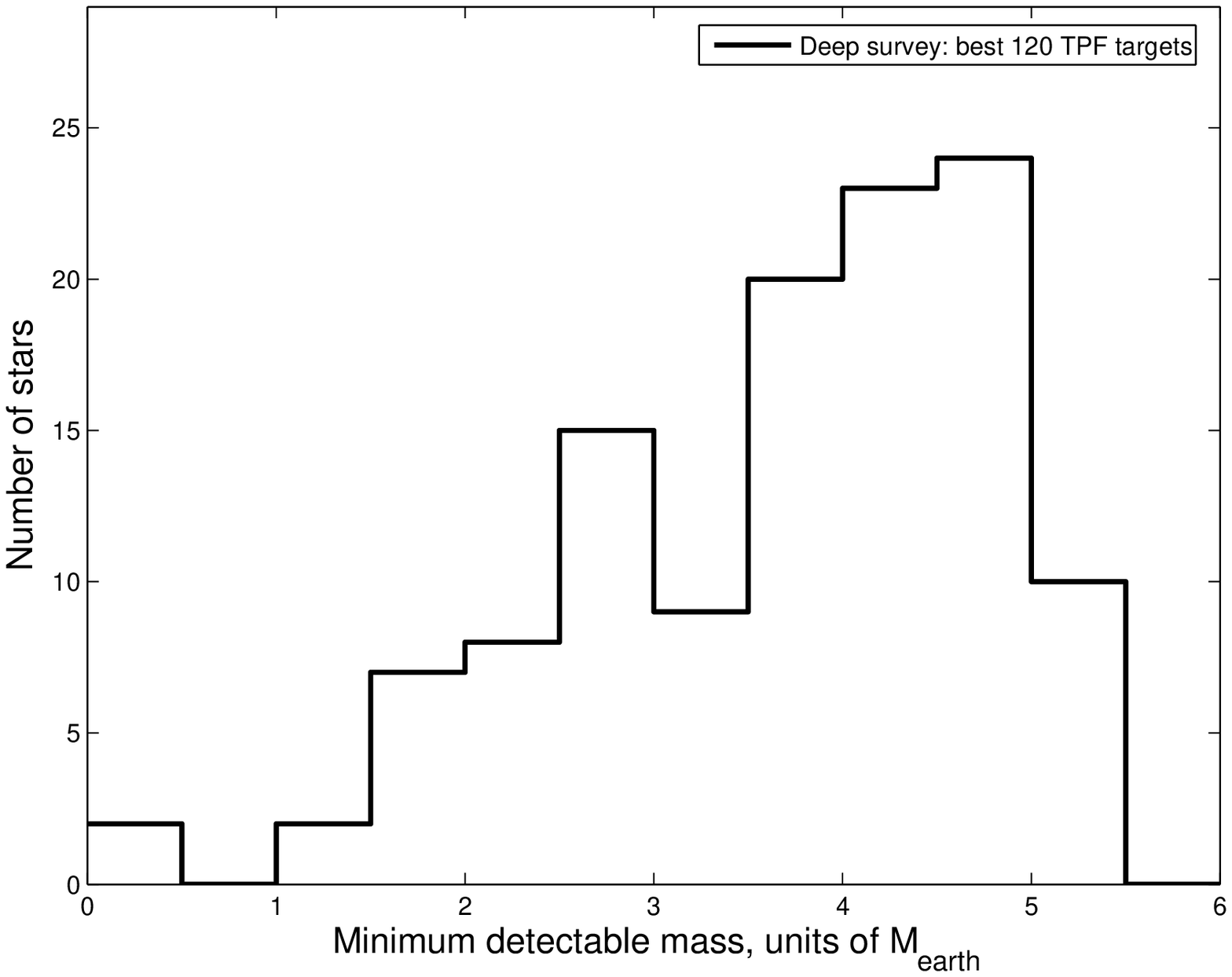}
\caption{Distribution of minimum detectable masses. Left: Deep
planet survey -- best 120 stars for SIM. Right: Deep planet survey
-- best 120 stars for TPF. For both plots, there are 104
two-dimensional measurements per star, with single measurement
precision $1.0~\mu as$, and minimum detectable mass is for 50\%
detection efficiency at detection threshold corresponding to 1\%
false-alarm probability. } \label{fig:8}
\end{figure}

\begin{figure}[p!]
\epsscale{1}\plottwo{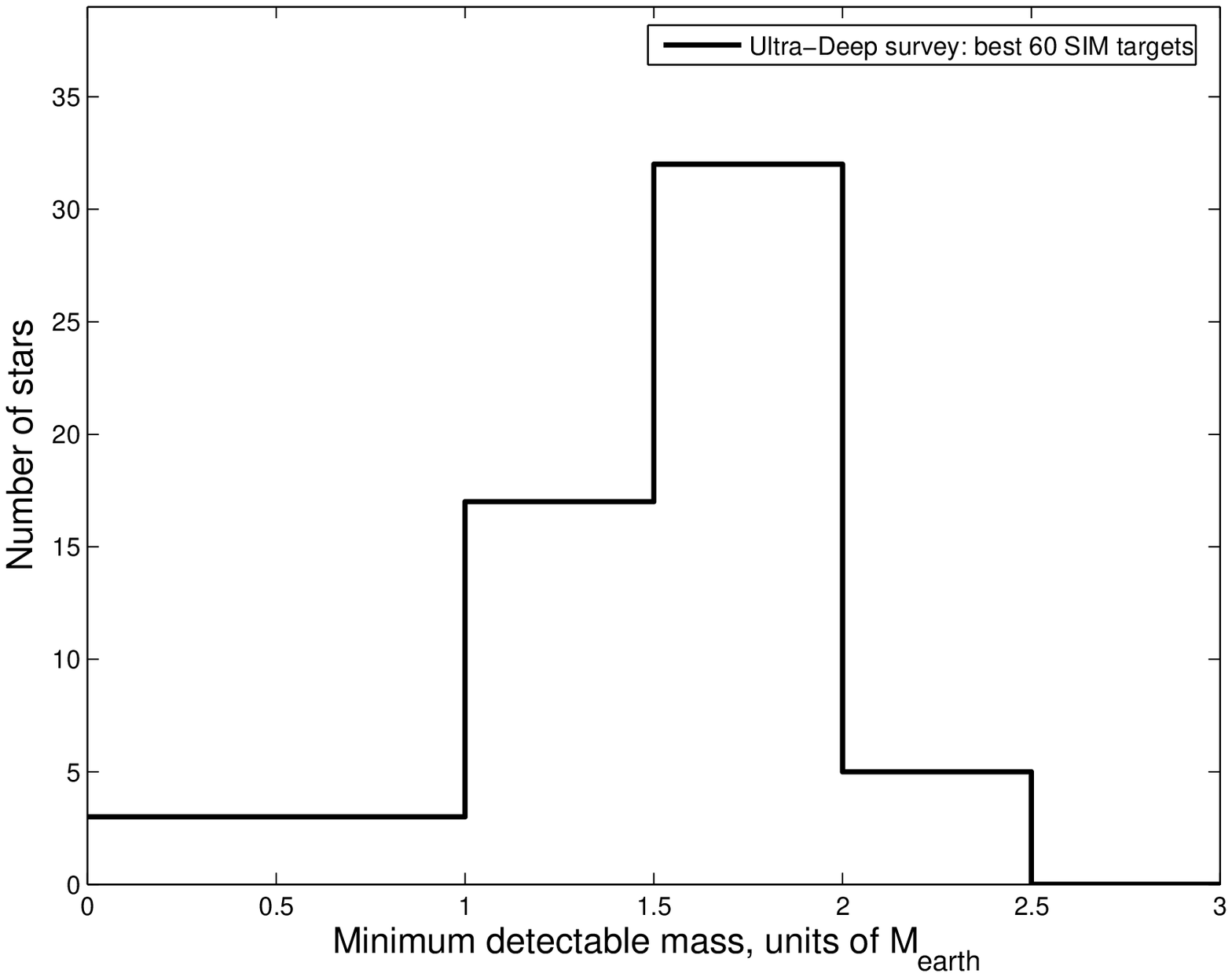}{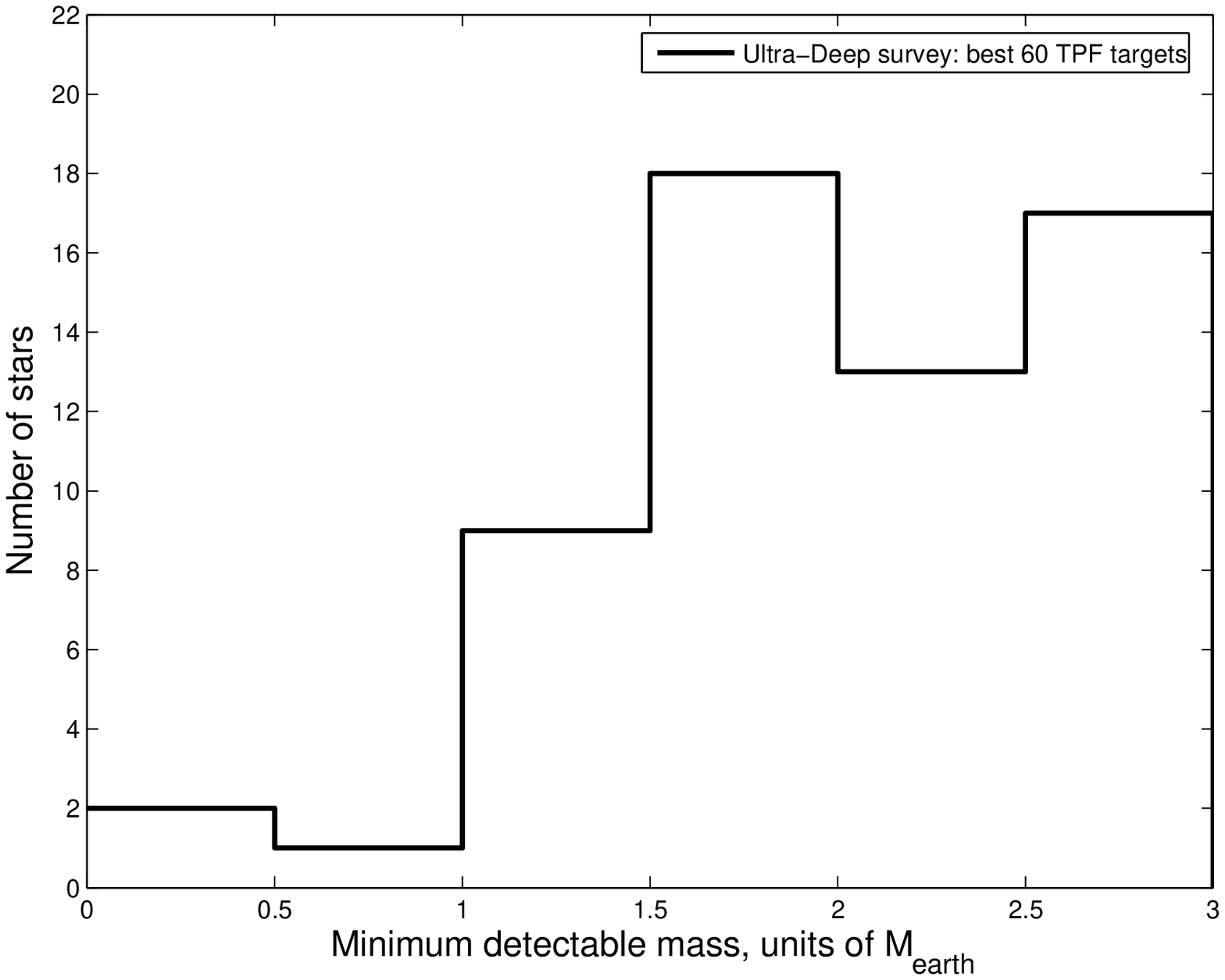}
\caption{Distribution of minimum detectable masses. Left: Ultra-Deep
planet survey -- best 60 stars for SIM. Right: Ultra-Deep planet
survey -- best 60 stars for TPF. For both plots, there are 208
two-dimensional measurements per star, with single measurement
precision $1.0~\mu as$, and minimum detectable mass is for 50\%
detection efficiency at detection threshold corresponding to 1\%
false-alarm probability.} \label{fig:9}
\end{figure}

\begin{figure}[p!]
\epsscale{1}\plottwo{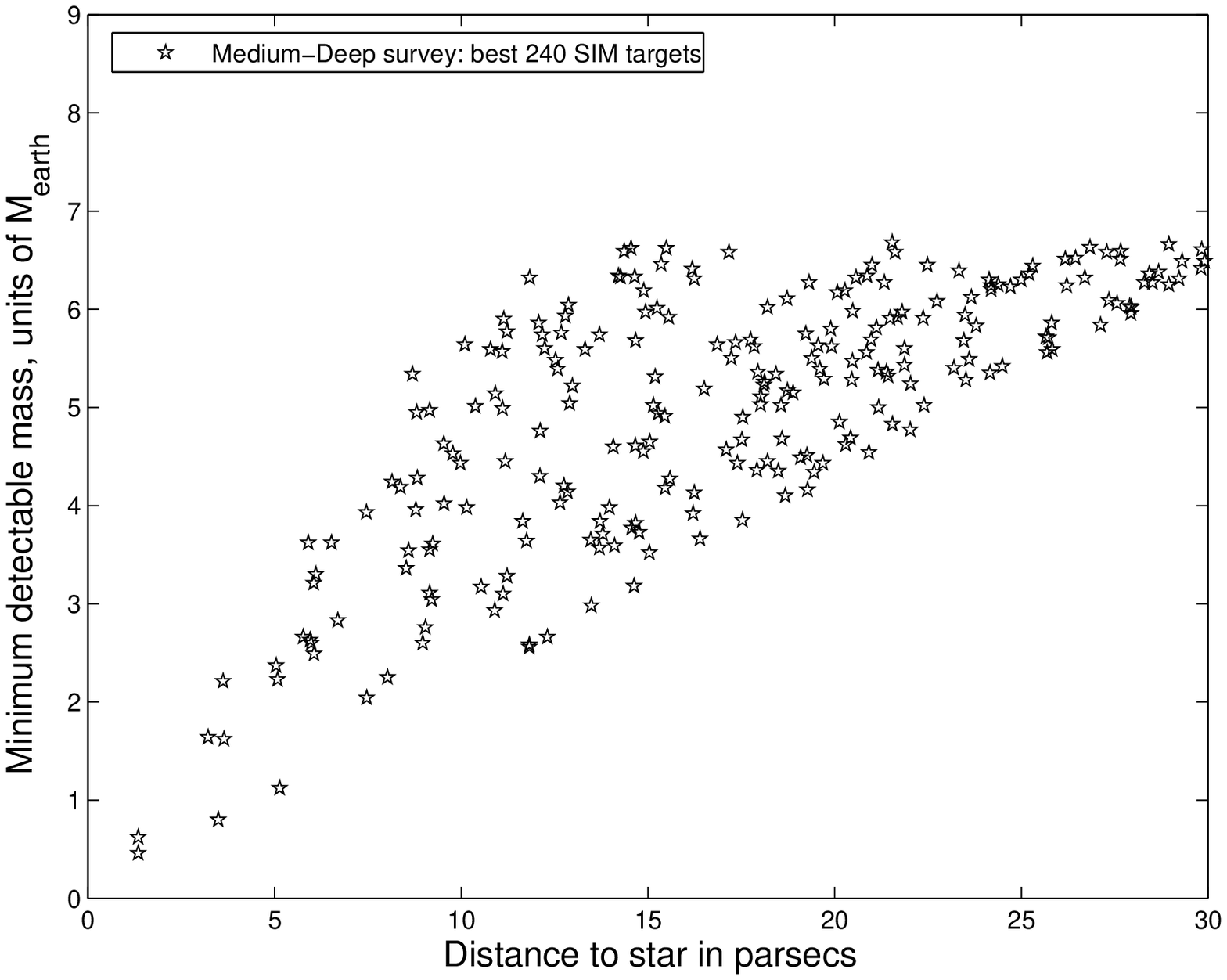}{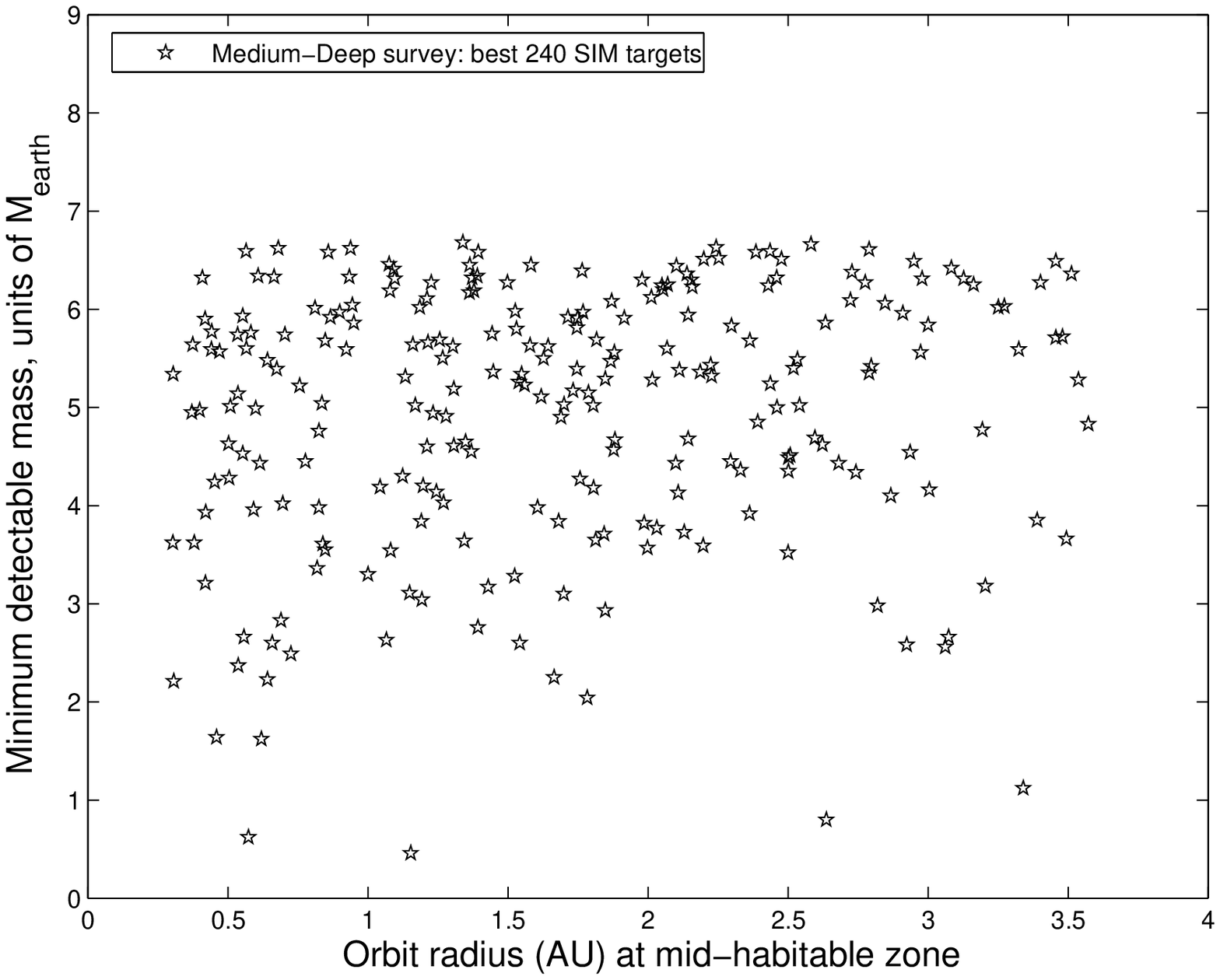} \caption{Left:
Sensitivity versus star distance. Right: Sensitivity versus
star-planet separation at mid-habitable zone. Medium-Deep planet
survey -- 240 stars, 52 two-dimensional measurements per star. For
both plots, single measurement precision is $1.0~\mu as$, and
minimum detectable mass is for 50\% detection efficiency at
detection threshold corresponding to 1\% false-alarm probability.}
\label{fig:7}
\end{figure}

\begin{figure}[p!]
\plottwo{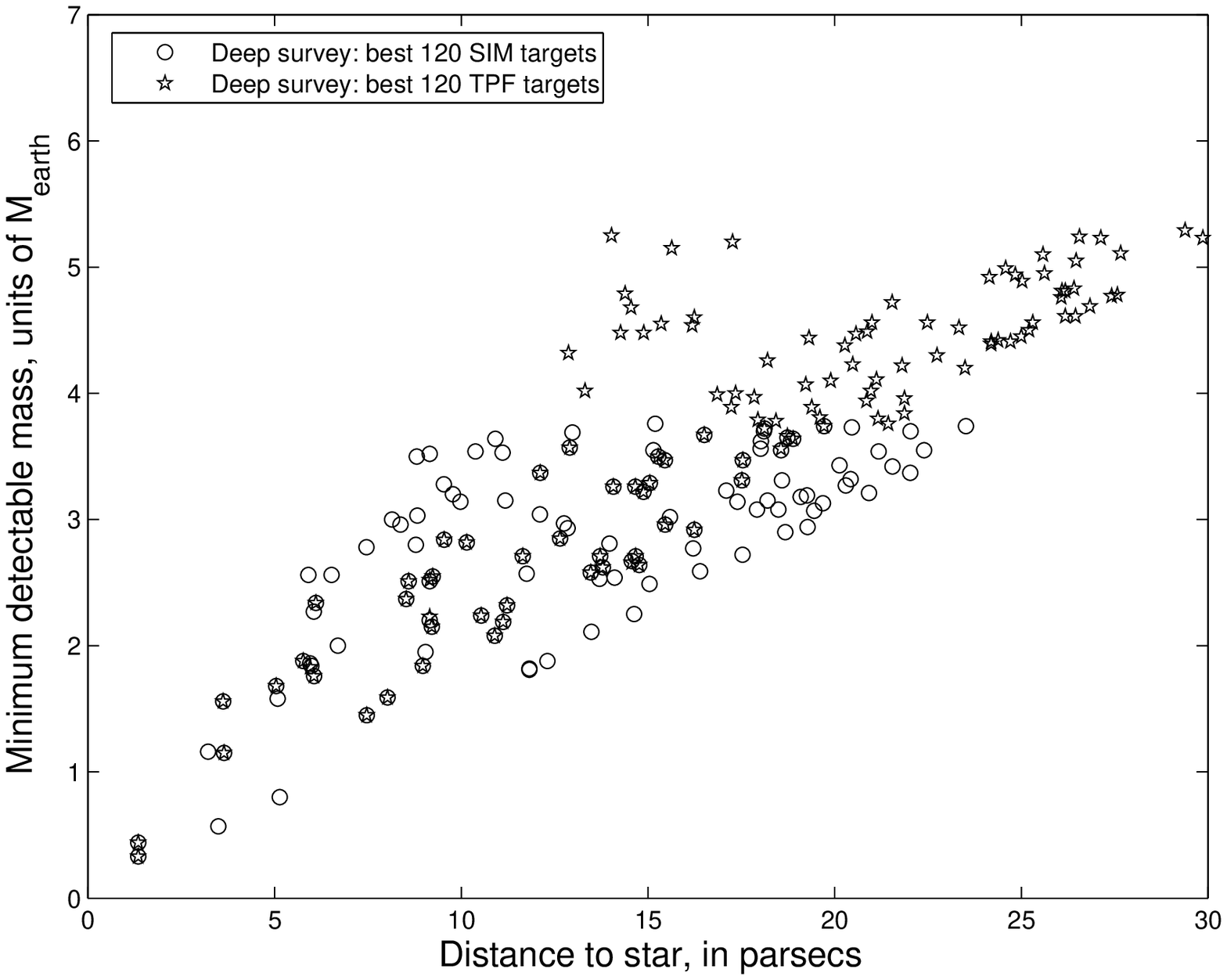}{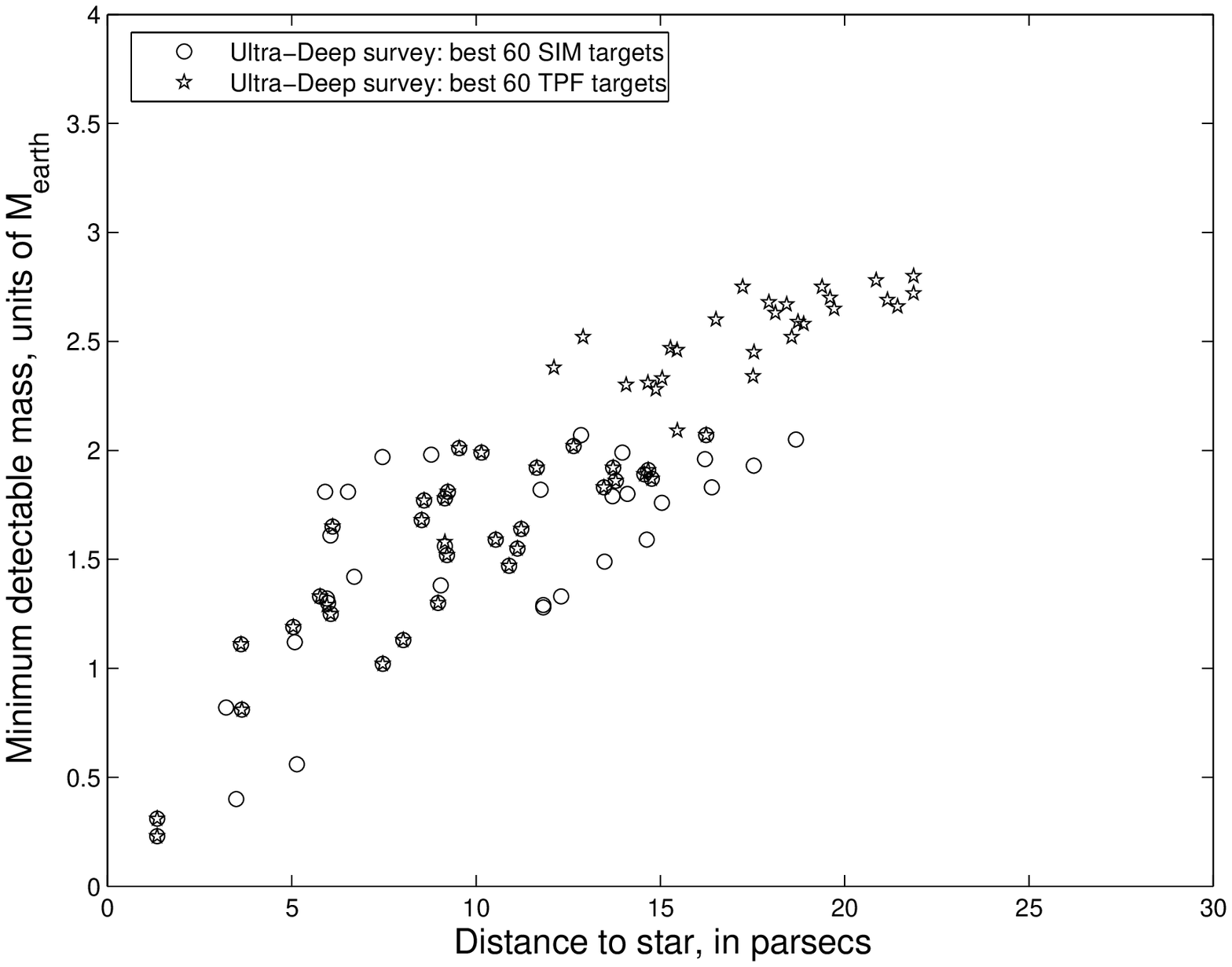} \caption{Detection
sensitivity versus star distance. Left: Deep planet survey -- best
SIM targets versus best TPF targets: 120 stars, 104 two-dimensional
measurements per star. Right: Ultra-Deep planet survey -- best SIM
targets versus best TPF targets: 60 stars, 208 two-dimensional
measurements per star. For both plots, single measurement precision
is $1.0~\mu as$, and minimum detectable mass is for 50\% detection
efficiency at detection threshold corresponding to 1\% false-alarm
probability.} \label{fig:10}
\end{figure}

\begin{figure}[p!]
\plottwo{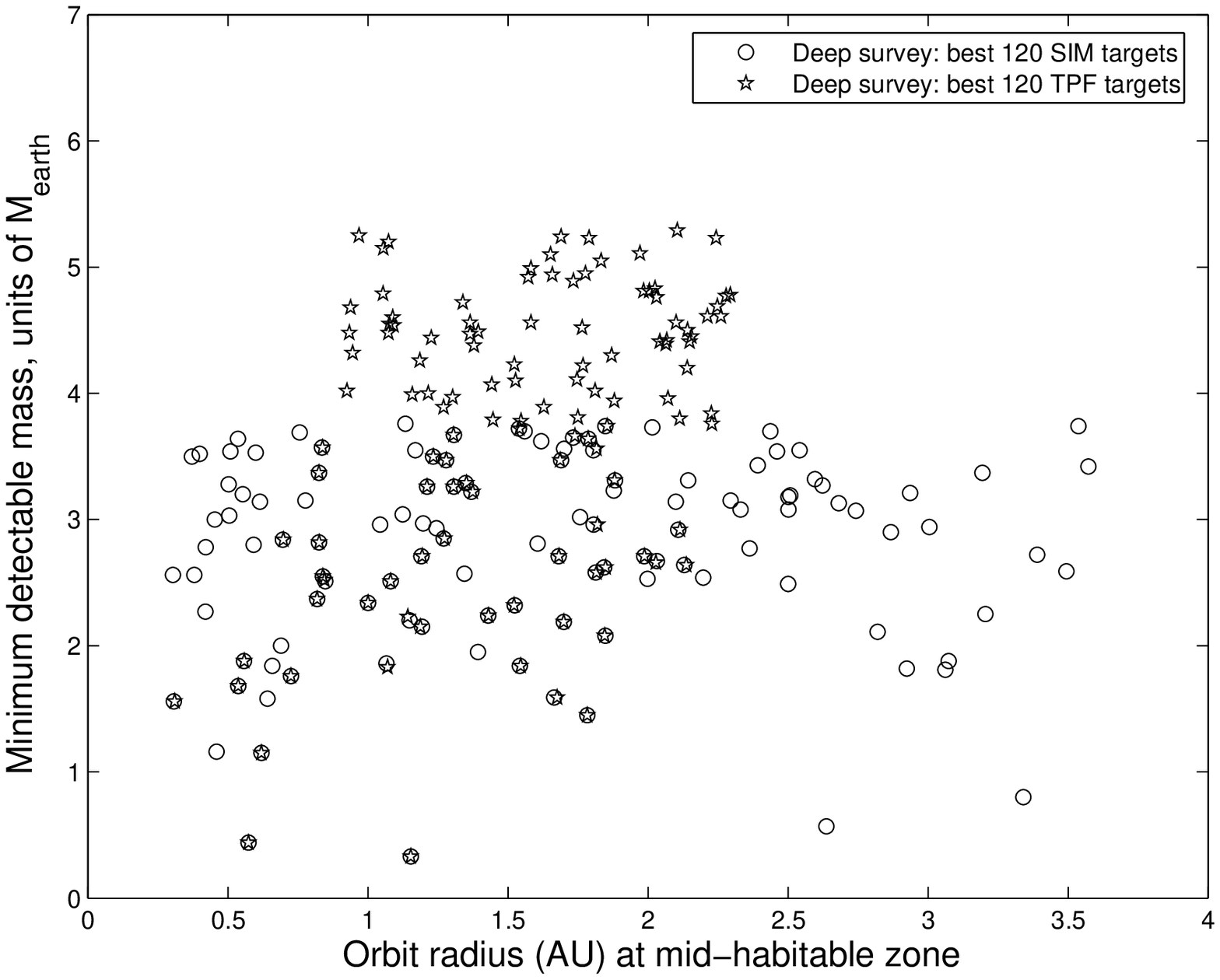}{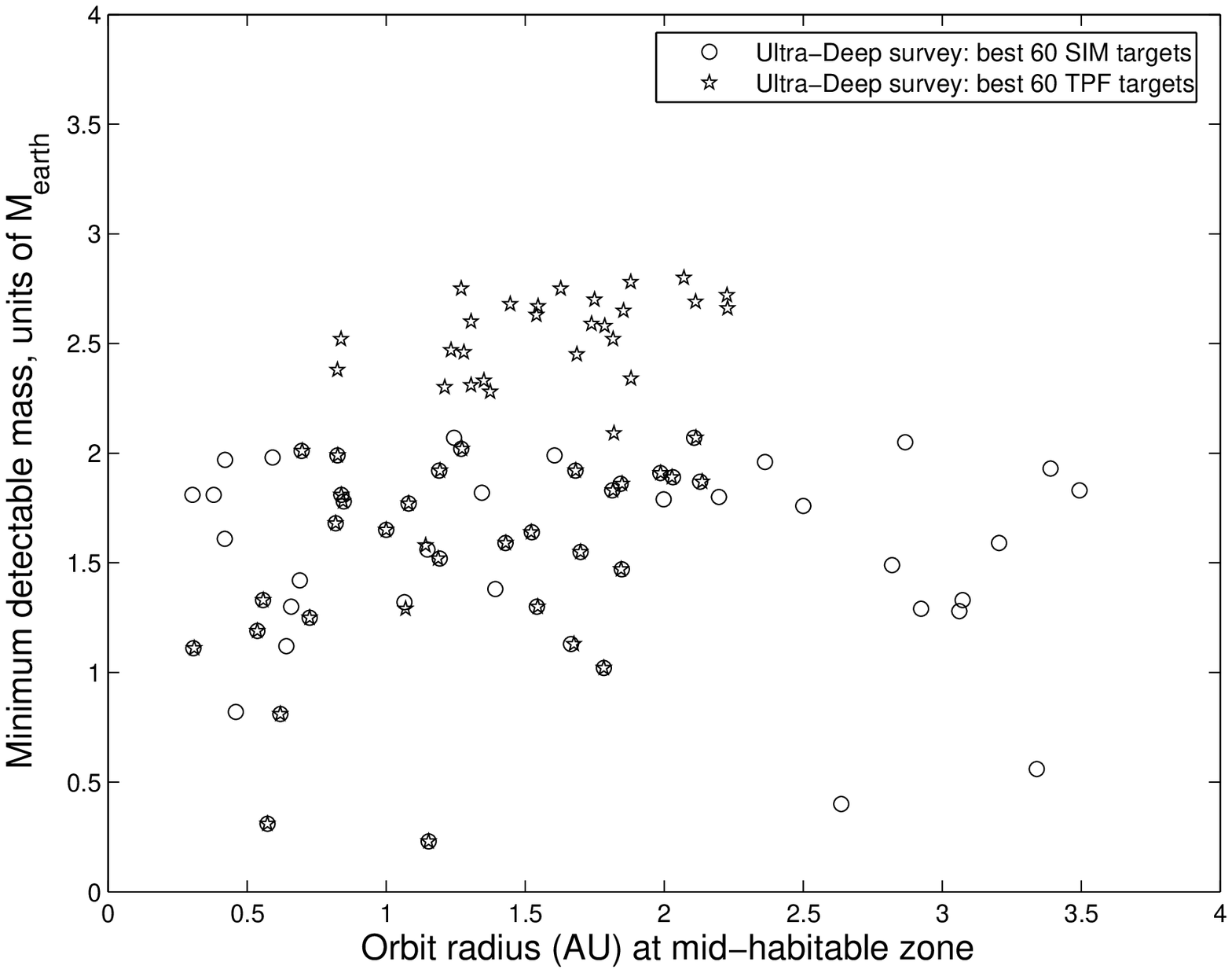} \caption{Detection
sensitivity versus star-planet separation at mid-habitable zone.
Left:  Deep planet survey -- best SIM targets versus best TPF
targets: 120 stars, 104 two-dimensional measurements per star.
Right: Ultra-Deep planet survey -- best SIM targets versus best TPF
targets: 60 stars, 208 two-dimensional measurements per star. For
both plots, single measurement precision is $1.0~\mu as$, and
minimum detectable mass is for 50\% detection efficiency at
detection threshold corresponding to 1\% false-alarm probability.}
\label{fig:11}
\end{figure}

\begin{figure}[p!] \plotone{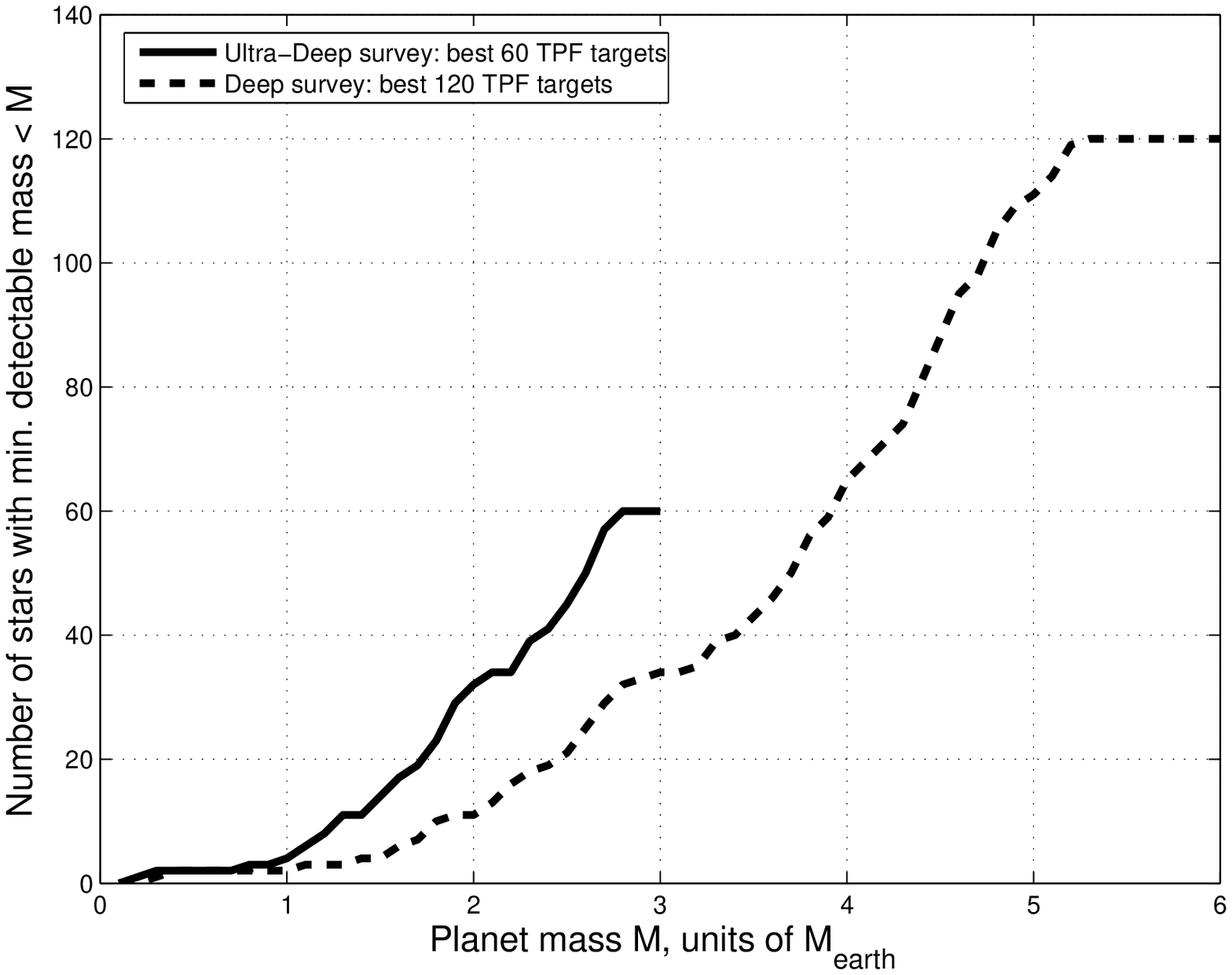}
\caption{SIM planet mass detection limits. Cumulative distribution
of minimum detectable mass, for surveys of the best TPF targets.
Single measurement precision is $1.0~\mu as$, and minimum detectable
mass is for 50\% detection efficiency at detection threshold
corresponding to 1\% false-alarm probability. These results are
\emph{independent} of assumptions regarding the mass distribution
and occurrence rate of terrestrial planets. }\label{fig:15}
\end{figure}

\begin{figure}[p!] \plotone{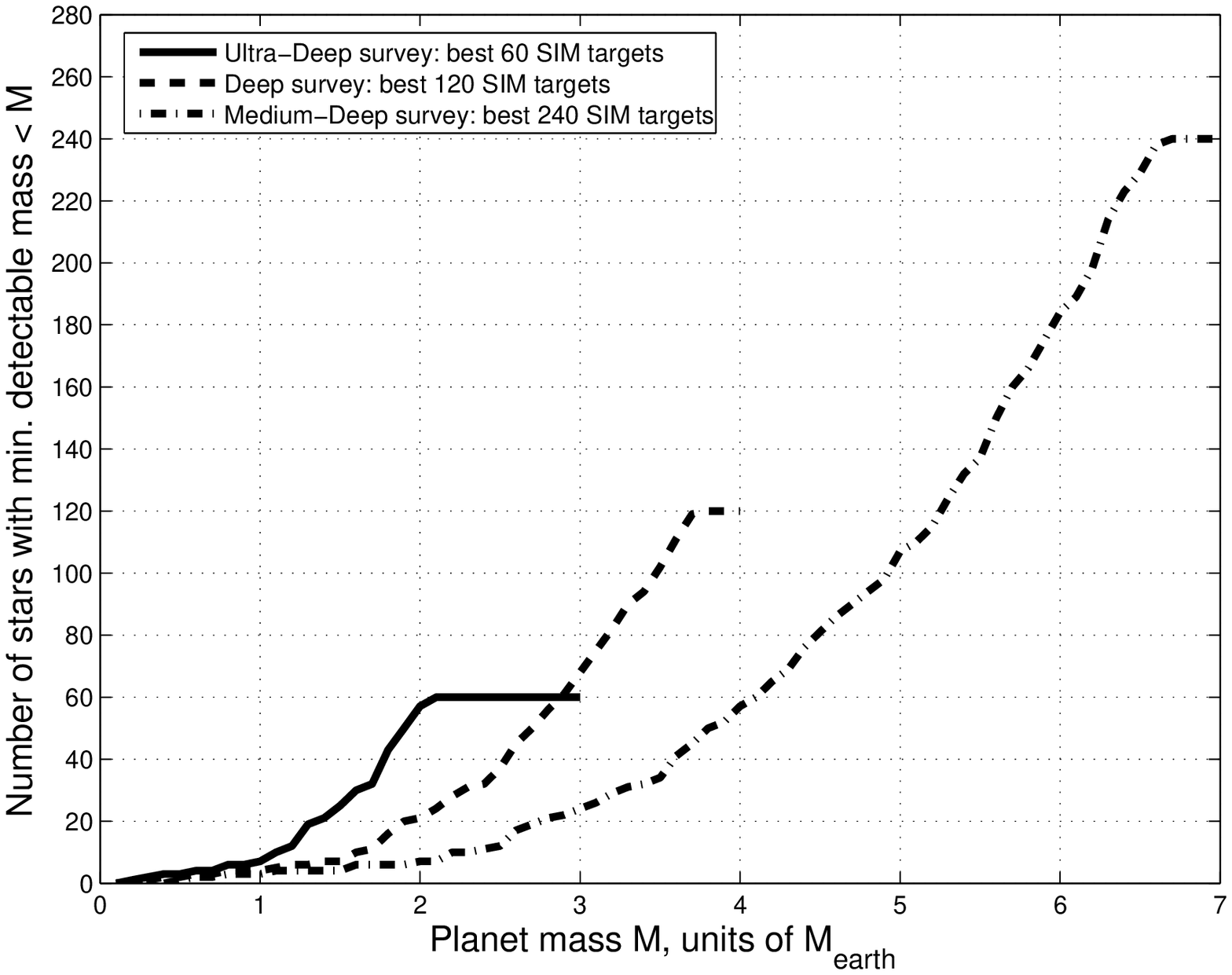}
\caption{SIM planet mass detection limits. Cumulative distribution
of minimum detectable mass, for surveys of the best SIM targets.
Single measurement precision is $1.0~\mu as$, and minimum detectable
mass is for 50\% detection efficiency at detection threshold
corresponding to 1\% false-alarm probability. These results are
\emph{independent} of assumptions regarding the mass distribution
and occurrence rate of terrestrial planets. }\label{fig:115}
\end{figure}

For the hypothetical SIM target list, Figures~\ref{fig:6},
\ref{fig:8}A and~\ref{fig:9}A show histograms of the minimum
detectable mass for the Medium-Deep, Deep and Ultra-Deep surveys of
the best SIM targets. Figures~\ref{fig:8}B and~\ref{fig:9}B show
distributions of minimum detectable masses for the Deep and
Ultra-Deep surveys for the best TPF-C targets. There is no
Medium-Deep survey for the TPF-C target list, since there are only
$\sim120$ stars meeting the requirements that the habitable zone lie
outside the inner working angle of 62 mas and close enough to the
star so that the contrast exceeds $10^{-10}$. The main results of
this study are presented in Figures~\ref{fig:15} and ~\ref{fig:115}.
These figures show the mass limits of SIM planet detection,
comparing cumulative distributions of minimum detectable mass for
surveys of the best TPF-C and SIM targets, respectively. These
results are summarized in Table ~\ref{tbl:1}. Figures~\ref{fig:7}A
and~\ref{fig:10}AB show minimum detectable mass versus stellar
distance. Figures~\ref{fig:7}B and~\ref{fig:11}AB show minimum
detectable mass versus star-planet separation at mid-habitable zone.
We find that SIM can probe the best 60, 120 and 240 planet search
targets down to planet masses of 2, 4, and 7 $M_{\earth}$,
respectively. It is important to note that these results depend only
on assumptions about the SIM instrument. They do not depend on the
astrophysics of planet mass distribution and occurrence frequency.

\subsection{Comparison with earlier studies}
Several previous studies \citep{S2002, S2003, FT2003} have addressed
SIM's detection and orbital characterization capabilities. These
studies employed the $\chi^2$ test rather than the periodogram for
planet detection. A detection is registered when the $\chi^2$ test
warrants a rejection of the null (no-planet) hypothesis. The studies
of Sozzetti et al. adopt a significance level of 95\% when
determining whether a planet has been detected. They define a scaled
signal S, as the ratio of the astrometric amplitude, $\alpha$, and
the single measurement astrometric accuracy, $\sigma_d$. They
determine that S must be greater than 2.2 to detect a planet with
95\% probability, and assume that SIM will make only 24 2-D
observations per star surveyed. Other points of difference between
their work and this work are: they assume that planetary orbits span
the full range of eccentricities; we focus only on planets in the
habitable zone; they assume a single measurement accuracy of $2~\mu
as$ (and higher, for faint stars).

Ford \& Tremaine (2003) generally accept the \citep{S2002}
conclusions regarding SIM's detection efficiency. They further
investigate the issue via similar Monte Carlo simulations for
planets around stars of spectral types F, G, K and M, within 100 pc
and up to a V magnitude of 10.5. They consider a range of single
measurement accuracies of 1, 1.4, and 2 $\mu as$ associated with
samples of 120, 240, and 480 stars. They also consider a two-tier
observing strategy with the first tier of stars measured to $1~\mu
as$ accuracy and the second tier measured to $4 ~\mu as$ accuracy;
they consider both five and ten year missions. As in this work, they
adopt the power-law planet mass distribution \citep{TT2002}
consistent with currently known planets discovered by the radial
velocity method.

We believe that the results of the present study represent a better
estimate of SIM's likely science return in this field, for two
reasons. First, we have realistically modeled the likely SIM
performance and observing scenario, taking total observation time
into consideration. Second, the joint periodogram represents a more
effective method of extracting planetary signals from astrometric
data than the $\chi^{2}$ test of the null hypothesis.

\section{Discovery of terrestrial planets by SIM} \label{sec:Results2}
In this section we address two further questions: What is the mass
distribution and number of planets SIM will discover? What is SIM's
completeness for detection of terrestrial planets in the habitable
zone, i.e., what fraction of the terrestrial planet discovery space
does SIM probe?

To answer these questions, additional (astrophysical) assumptions
are needed. We first extrapolate the mass distribution of currently
known planets \citep{TT2002} to terrestrial planet masses. Next, for
each target list and survey strategy, we use the universal detection
efficiency curve to determine the detection efficiency versus planet
mass for each target star. These two relations allow us to determine
the expected mass distribution of terrestrial planets SIM will
detect. This can be compared with the expected mass distribution of
existing planets. In each mass bin, the completeness is defined to
be the ratio of the number of detected planets to the number of
expected planets. With an assumed value of $\eta_{terrestrial}$
(defined as the fraction of F, G, and K stars having terrestrial
planets in their habitable zones) we can also estimate the expected
\emph{number} of terrestrial planets SIM will discover in each mass
bin. Detailed results in the form of tables and plots are presented
in the next two subsections.

\subsection{What is the
mass distribution and number of planets SIM will discover?} Our
starting point is the universal detection efficiency versus SNR
curve for each star in the hypothetical target lists, as discussed
in the previous section. From Equation~\ref{eq:3}, $\alpha$ depends
on distance, stellar mass, planet mass, and luminosity (through size
of the habitable zone). Given these parameters, together with values
for $N_{obs}$ and $\sigma$, we can derive a relation between
detection efficiency and planet mass for each star in the target
list. To proceed further, we evaluate the hypothetical case that
each star has one terrestrial planet at its mid-habitable zone, with
mass drawn at random from the distribution $dN/dM
\varpropto~M^{-1.1}$ \citep{TT2002, M2005A}.

The expected probability distribution function for a detected planet
at a given mass is the product of detection efficiency and dN/dM, at
that mass. The probability distribution function has a peak, since
it is the product of the monotonically decreasing planet mass
distribution and the monotonically increasing, S-shaped detection
efficiency versus mass curve.

\begin{figure}[p!]
\plottwo{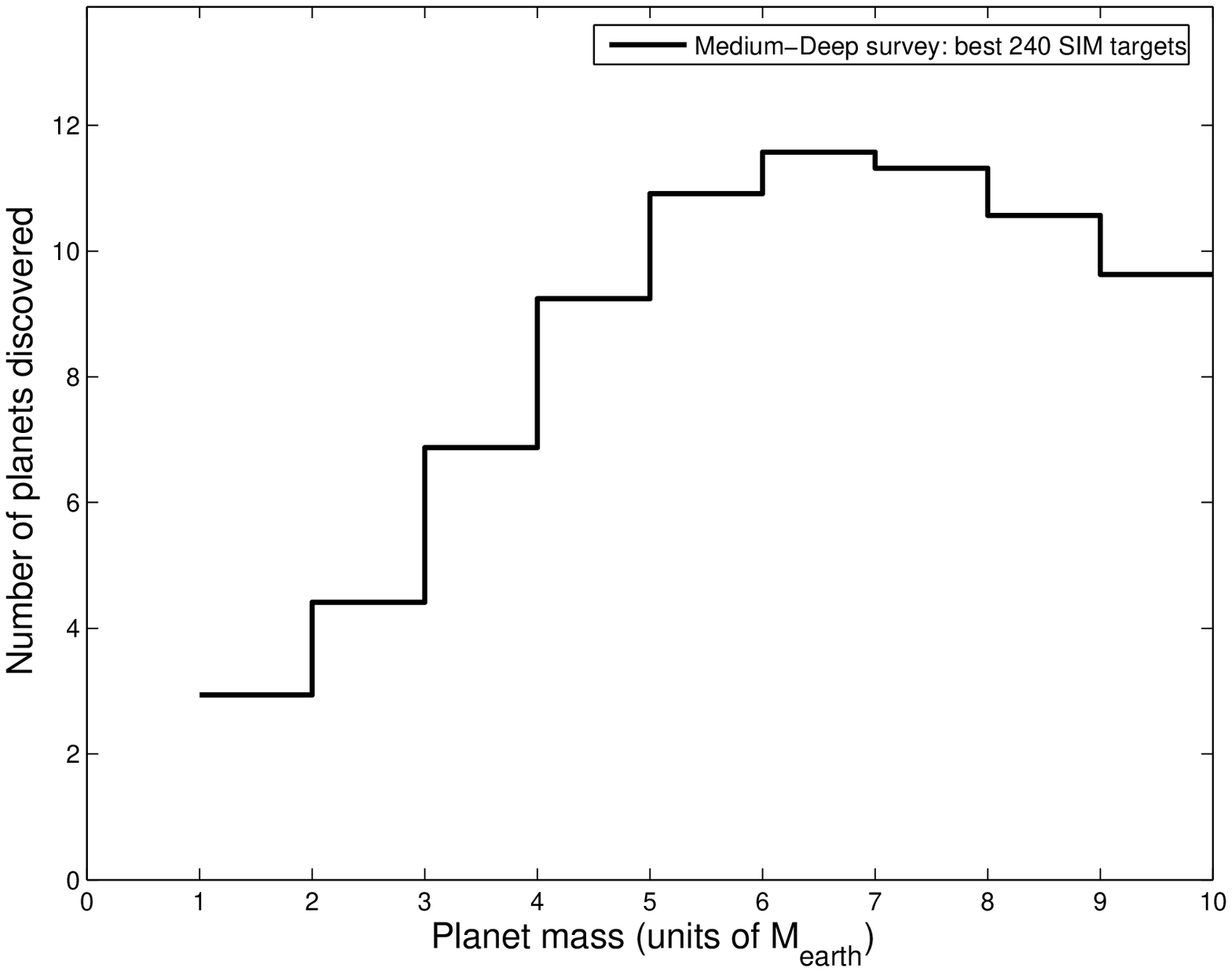}{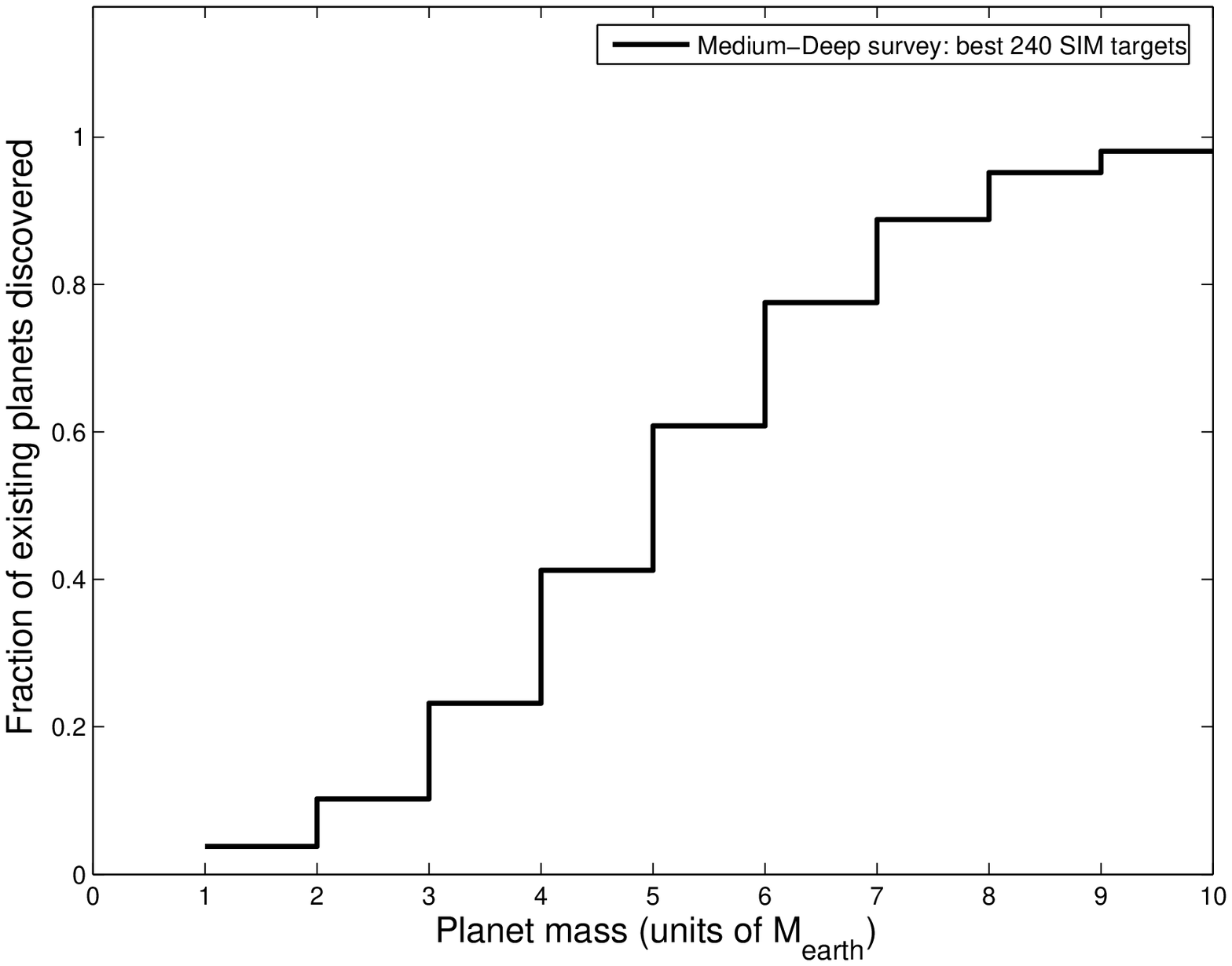} \caption{Left: Mass
distribution of SIM planet discoveries. Right: Completeness -- ratio
of number of detected planets to number of expected planets, in each
mass bin. Both plots are for Medium-Deep planet survey of best 240
SIM targets -- 52 two-dimensional measurements per star. Single
measurement precision is $1.0~\mu as$; detection threshold
corresponds to 1\% false-alarm probability; and planet mass
distribution is assumed $\varpropto~M^{-1.1}$.} \label{fig:12}
\end{figure}

\begin{figure}[p!]
\plottwo{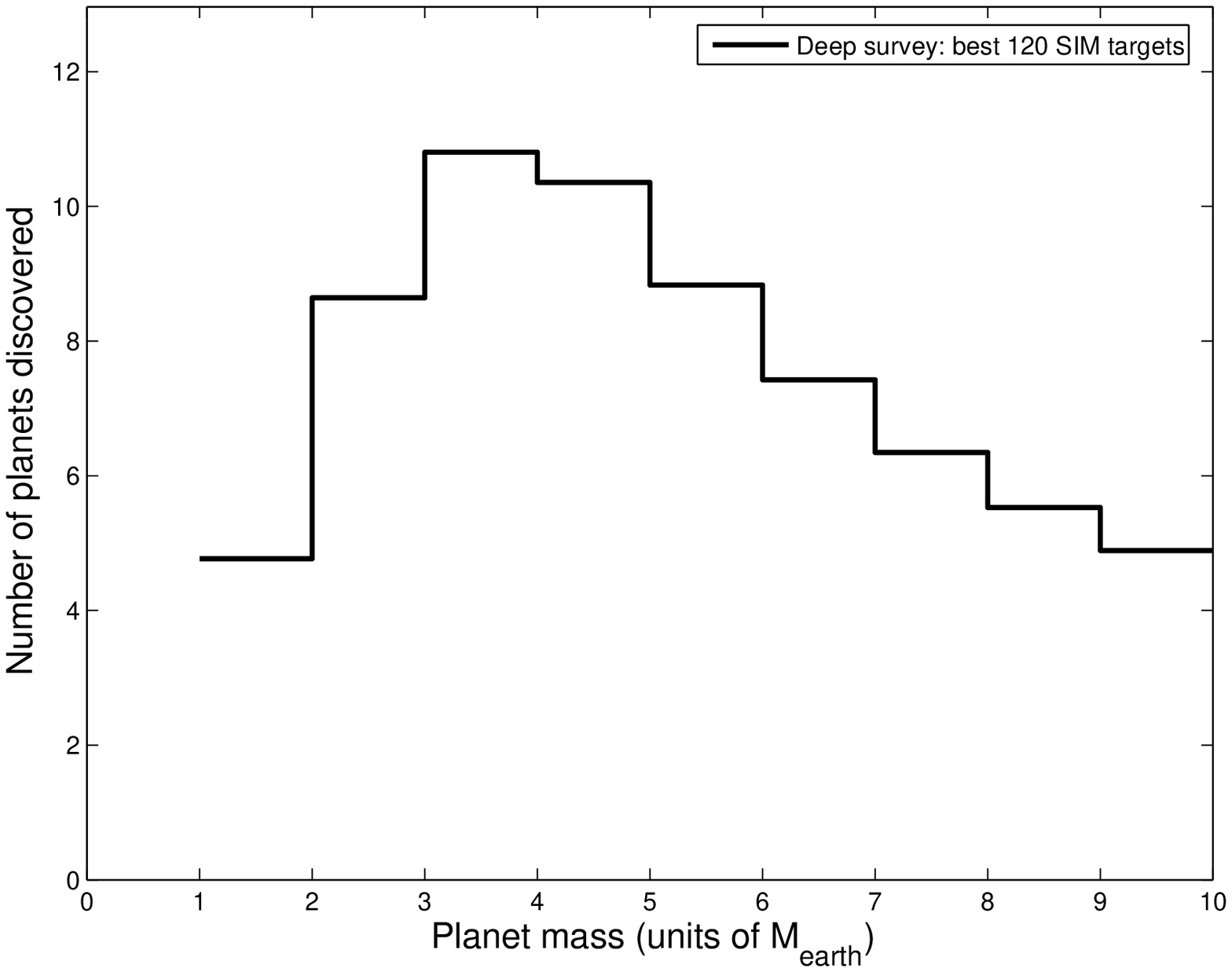}{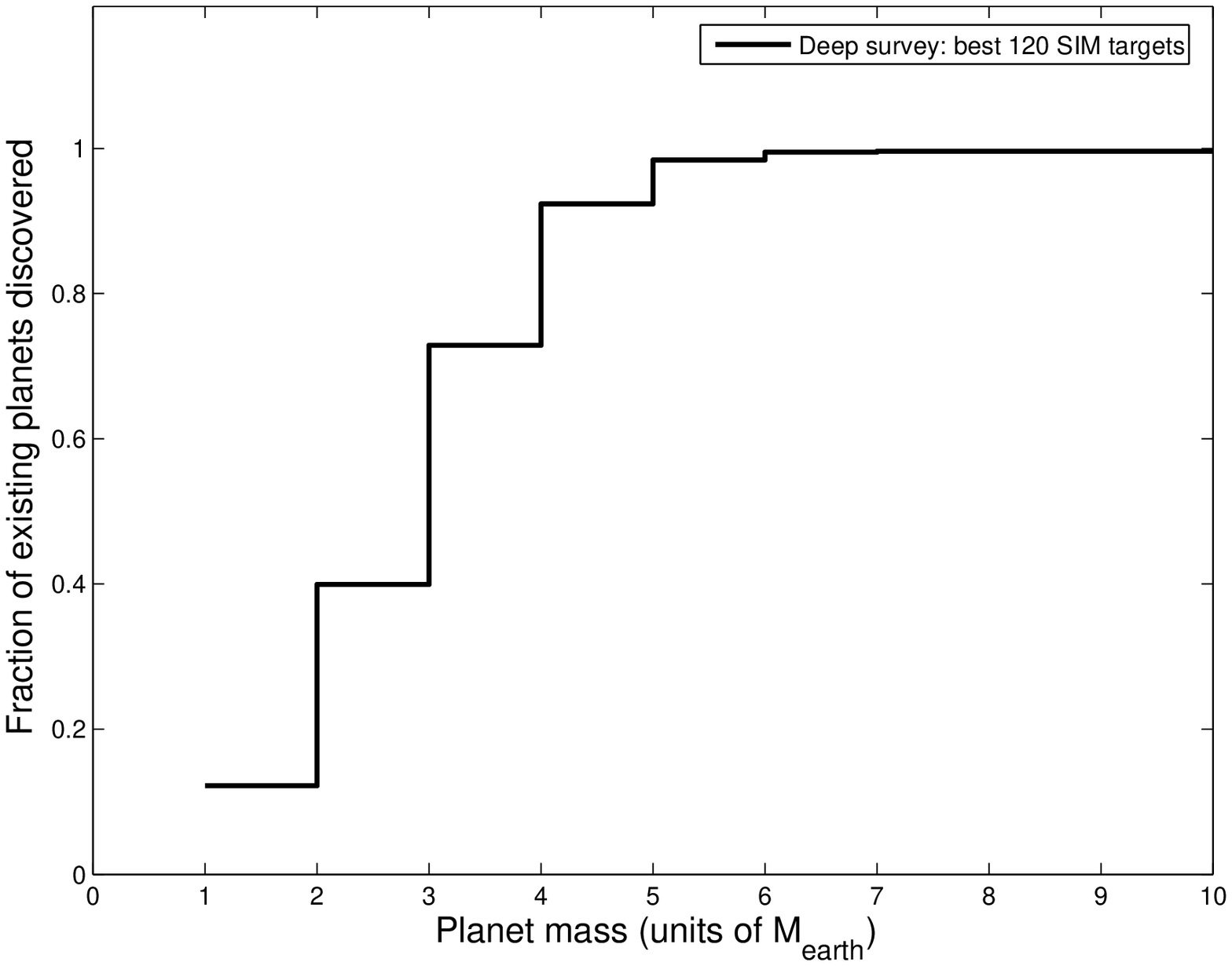} \caption{Left: Mass
distribution of SIM planet discoveries. Right: Completeness -- ratio
of number of detected planets to number of expected planets, in each
mass bin. Both plots are for Deep planet survey of best 120 SIM
targets -- 104 two-dimensional measurements per star. Single
measurement precision is $1.0~\mu as$; detection threshold
corresponds to 1\% false-alarm probability; and planet mass
distribution is assumed $\varpropto~M^{-1.1}$.} \label{fig:113}
\end{figure}

\begin{figure}[p!]
\plottwo{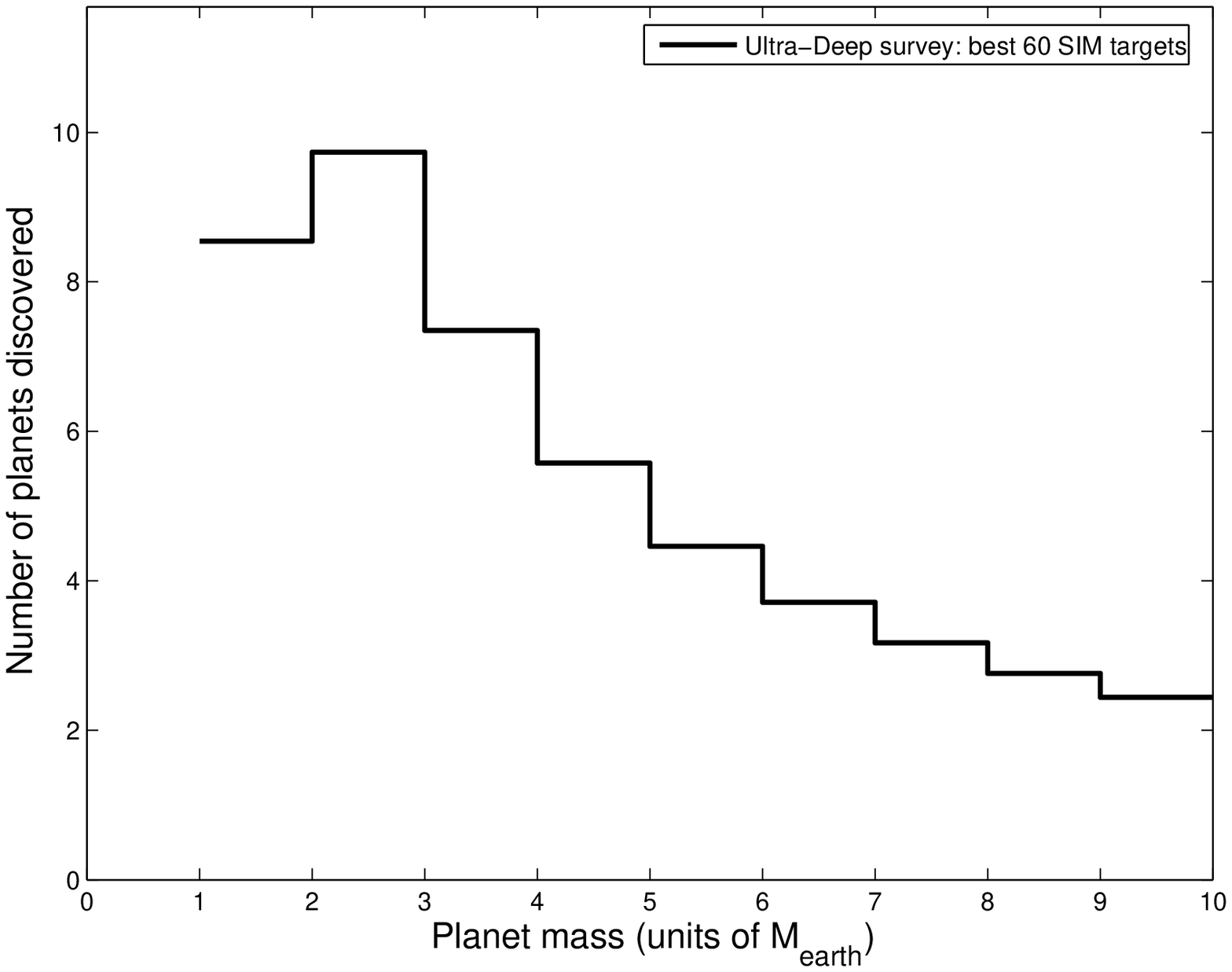}{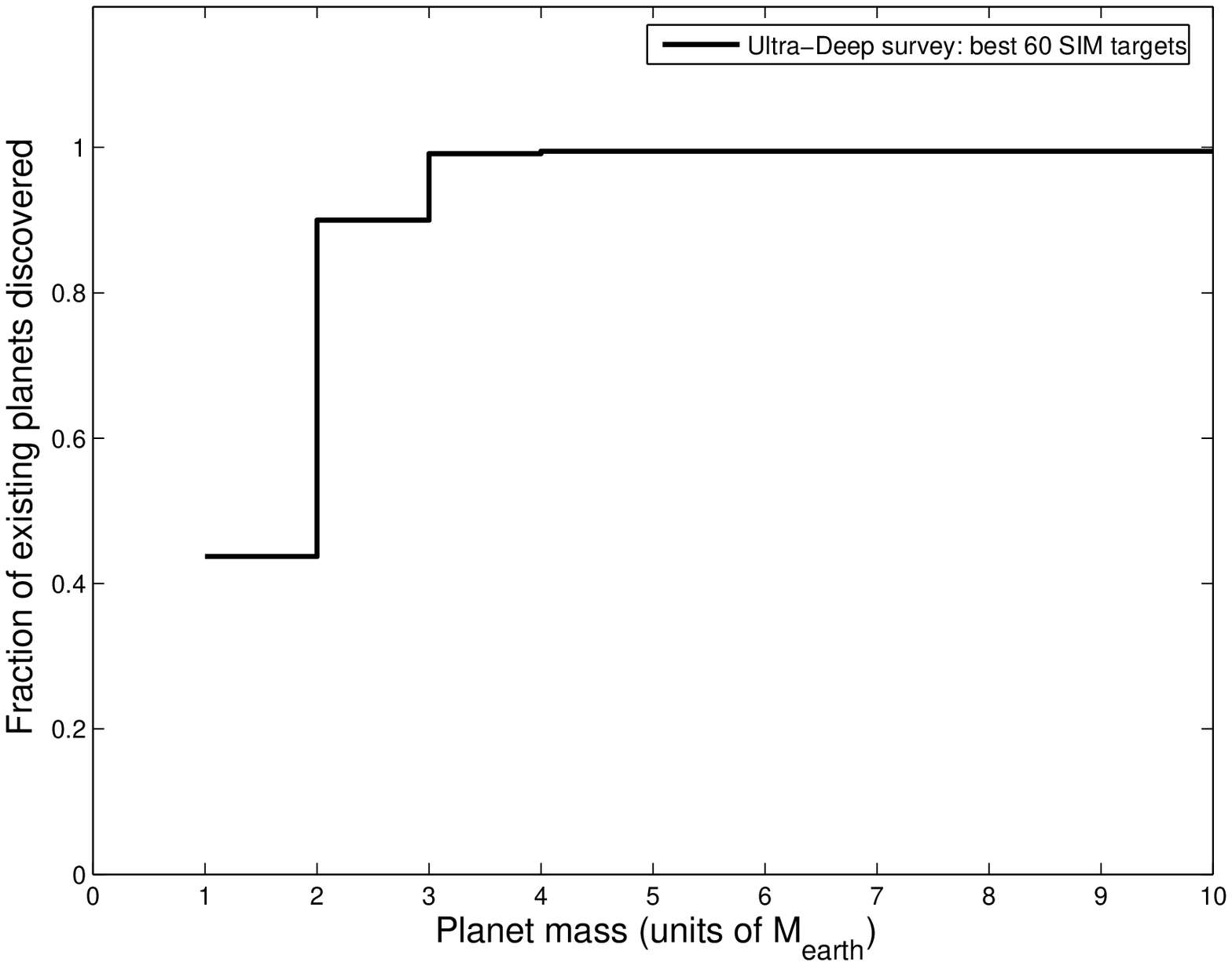} \caption{Left: Mass
distribution of SIM planet discoveries. Right: Completeness -- ratio
of number of detected planets to number of expected planets, in each
mass bin. Both plots are for Ultra-Deep planet survey of best 60 SIM
targets -- 208 two-dimensional measurements per star. Single
measurement precision is $1.0~\mu as$; detection threshold
corresponds to 1\% false-alarm probability; and planet mass
distribution is assumed $\varpropto~M^{-1.1}$.} \label{fig:114}
\end{figure}

\begin{figure}[p!]
\plottwo{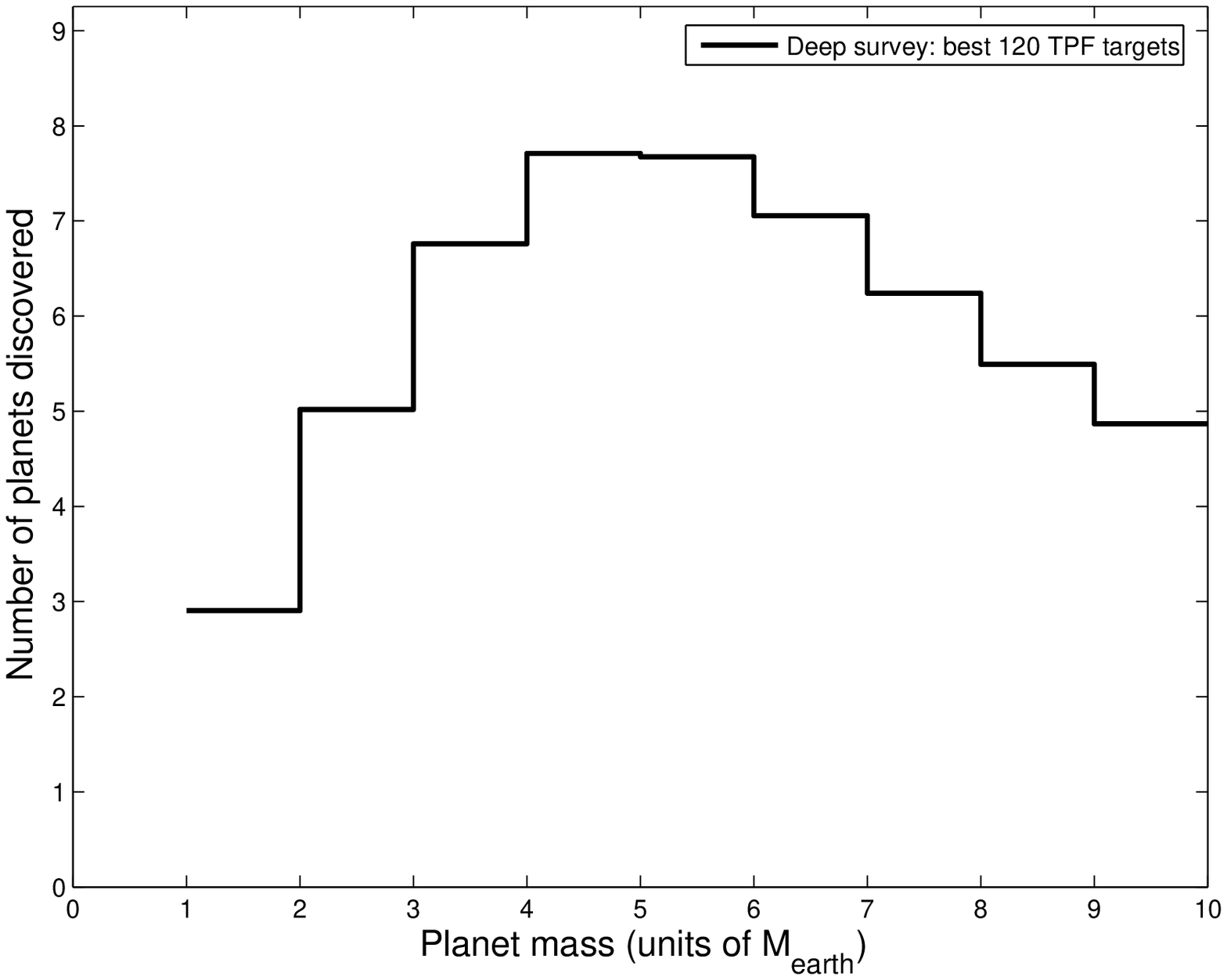}{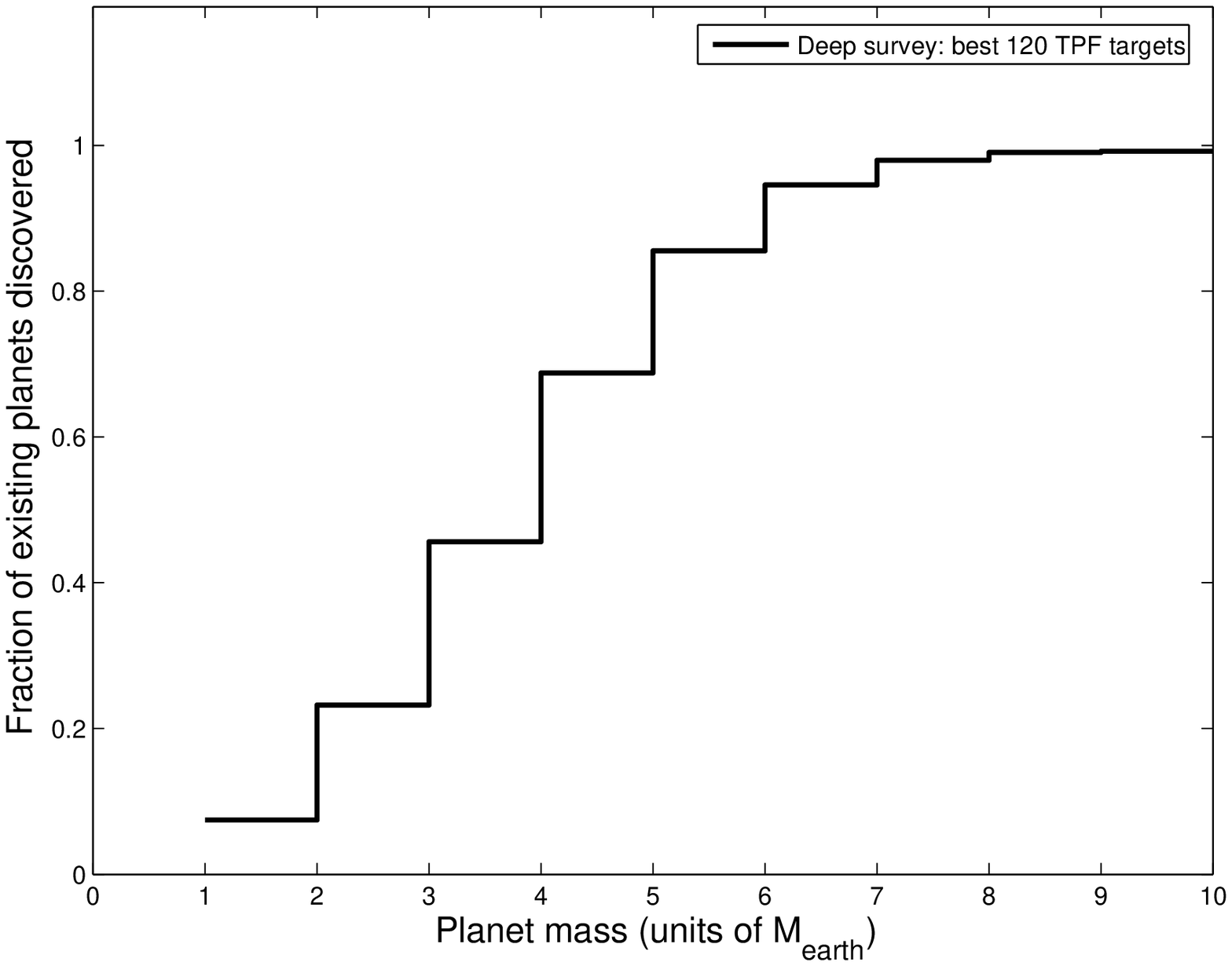} \caption{Left: Mass
distribution of SIM planet discoveries. Right: Completeness -- ratio
of number of detected planets to number of expected planets, in each
mass bin. Both plots are for Deep planet survey of best 120 TPF
targets -- 104 two-dimensional measurements per star. Single
measurement precision is $1.0~\mu as$; detection threshold
corresponds to 1\% false-alarm probability; and planet mass
distribution is assumed $\varpropto~M^{-1.1}$.} \label{fig:213}
\end{figure}

\begin{figure}[p!]
\plottwo{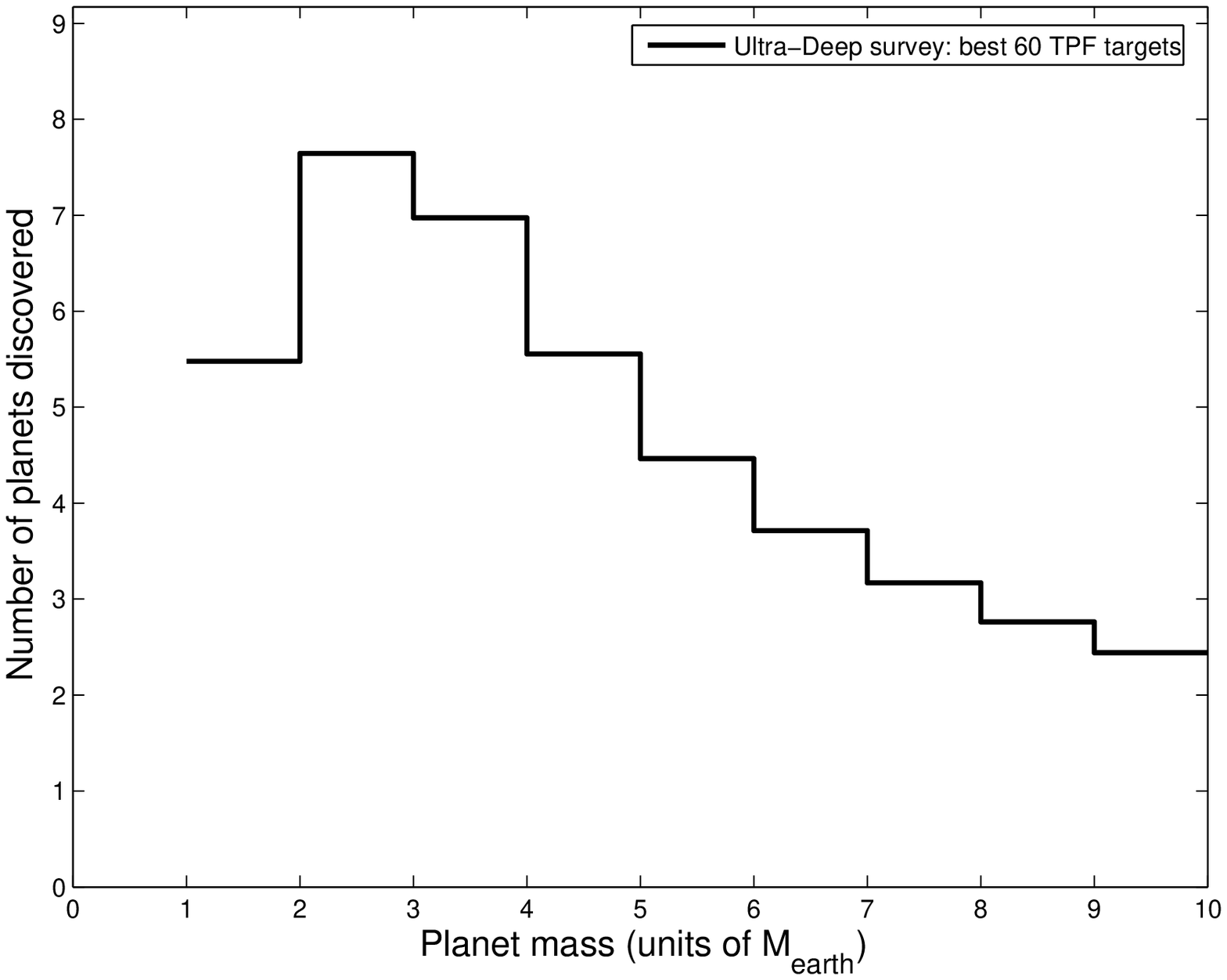}{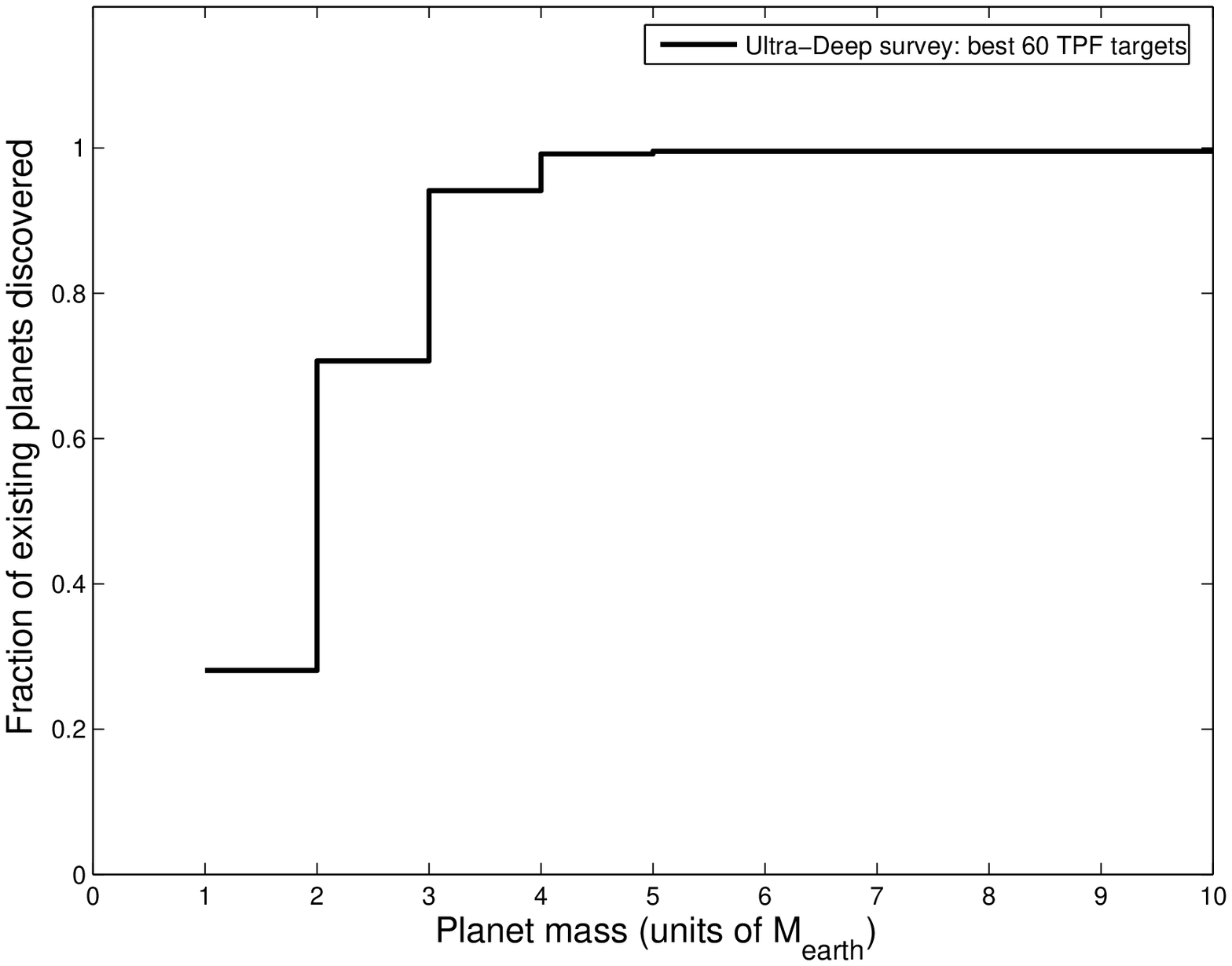} \caption{Left: Mass
distribution of SIM planet discoveries. Right: Completeness -- ratio
of number of detected planets to number of expected planets, in each
mass bin. Both plots are for Ultra-Deep planet survey of best 60 TPF
targets -- 208 two-dimensional measurements per star. Single
measurement precision is $1.0~\mu as$; detection threshold
corresponds to 1\% false-alarm probability; and planet mass
distribution is assumed $\varpropto~M^{-1.1}$.} \label{fig:214}
\end{figure}

Summing the expected probability distribution functions for all the
stars in the target list gives the distribution of the number of
terrestrial planets discovered per unit mass interval. Integrating
over unit mass bins from 1 to 10 $M_{\earth}$ gives a histogram of
the number of planets discovered in each mass bin.  Results for the
three survey modes for the SIM target list are shown in Figures
~\ref{fig:12}A, ~\ref{fig:113}A, and  ~\ref{fig:114}A; corresponding
results for surveys involving the TPF-C target list are shown in
Figures~\ref{fig:213}A and ~\ref{fig:214}A. The results are
summarized in Table ~\ref{tbl:2}.  These results are for the case of
$\eta_{terrestrial} = 1$, and are easily scaled to any other value
of $\eta_{terrestrial}$.

\begin{figure}[p!]
\plotone{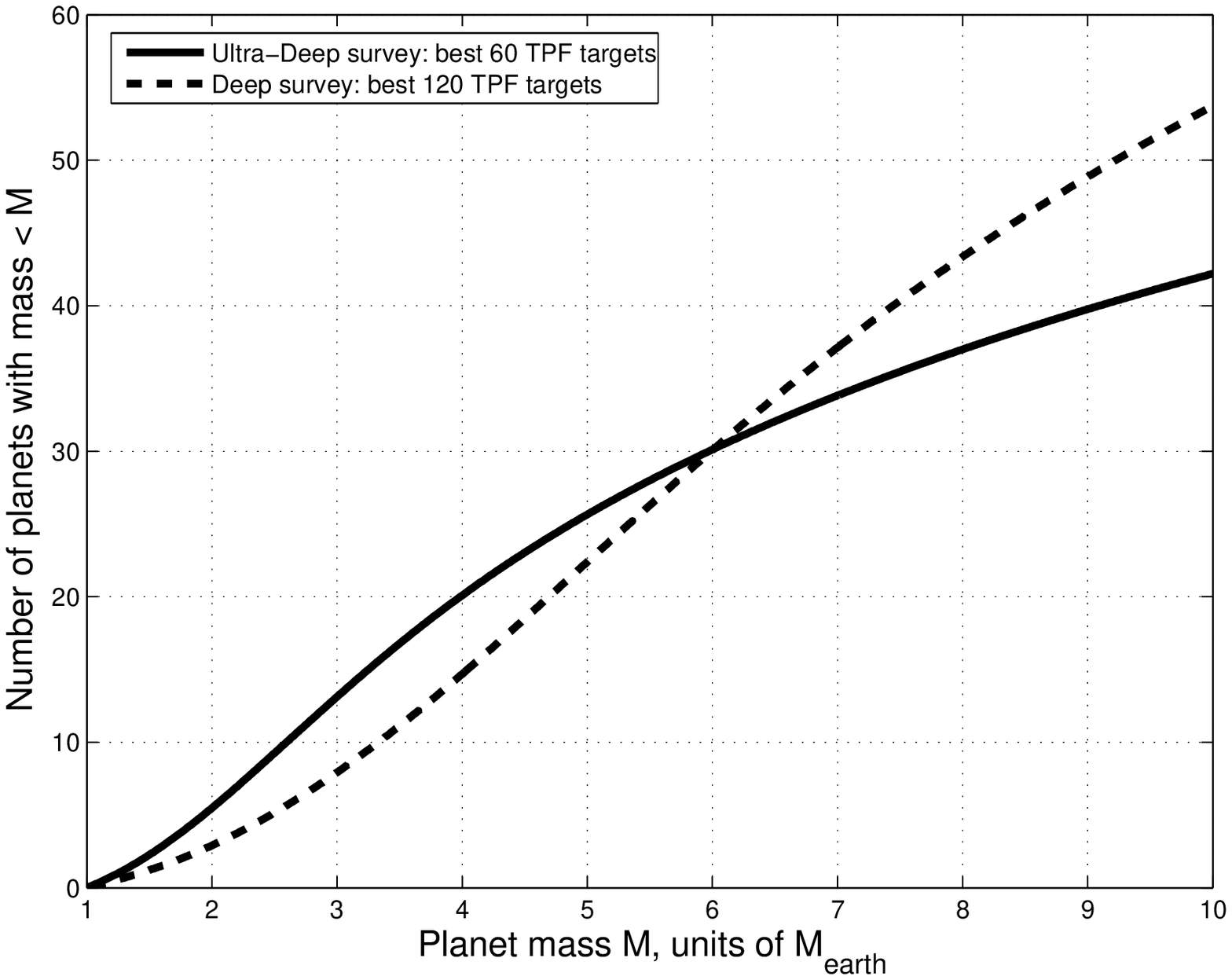} \caption{SIM planet discoveries. Cumulative
distribution of detected planet masses, for surveys of the best TPF
targets. Single measurement precision is $1.0~\mu as$, and detection
threshold corresponds to 1\% false-alarm probability. Assumes that
every target star has one terrestrial planet and that the planet
masses are distributed as $M^{-1.1}$.} \label{fig:16}
\end{figure}

\begin{figure}[p!]
\plotone{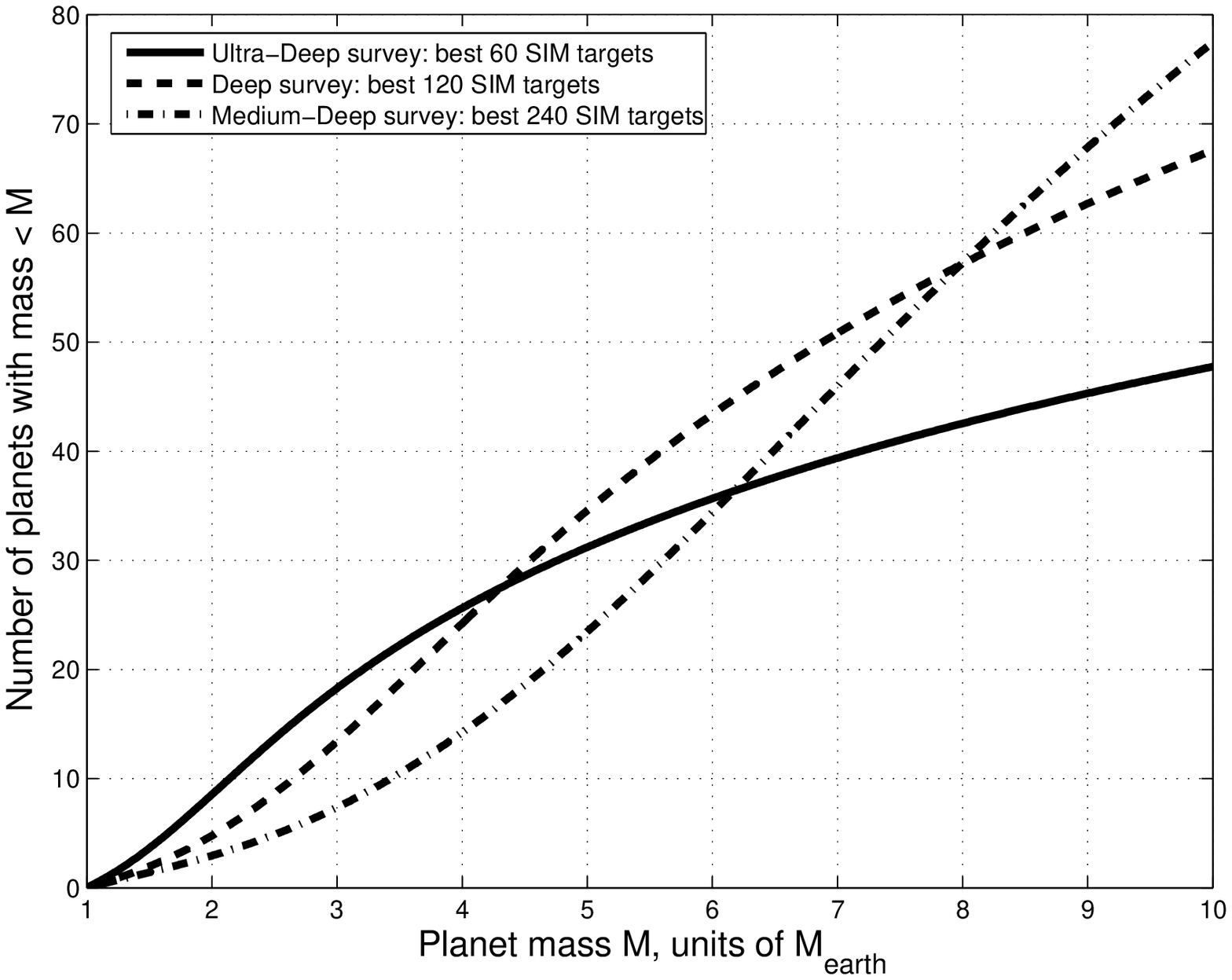} \caption{SIM planet discoveries. Cumulative
distribution of detected planet masses, for surveys of the best SIM
targets. Single measurement precision is $1.0~\mu as$, and detection
threshold corresponds to 1\% false-alarm probability. Assumes that
every target star has one terrestrial planet and that the planet
masses are distributed as $M^{-1.1}$.} \label{fig:116}
\end{figure}

Integrating the distribution of number of terrestrial planets
discovered per unit mass interval up to mass M gives the cumulative
distribution for the total number of planets discovered up to mass
M. Figures~\ref{fig:16} and ~\ref{fig:116} show results for surveys
of the best TPF-C and SIM targets, respectively. These plots reveal
the relative merits of the three survey strategies. For both target
lists, the Ultra-Deep survey nets more low-mass terrestrial planets,
but fewer total discoveries. For the SIM target list, the
Medium-Deep survey finds the most planets, but with the distribution
skewed toward higher masses.

\subsection{What fraction of the terrestrial planet discovery space
does SIM probe?}

For each mass bin, we determine completeness, which is the ratio of
detected planets to the number of expected planets. We also
determine cumulative completeness, the ratio of the number of
detected planets below mass M to the number of expected planets
below mass M. Completeness depends on our assumption that the
probability distribution function (pdf) of terrestrial planet masses
is $\varpropto~M^{-1.1}$; however no assumption about
$\eta_{terrestrial}$ is needed.
Figures~\ref{fig:12}B,~\ref{fig:113}B, and~\ref{fig:114}B show
completeness for Medium-Deep, Deep, and Ultra-Deep surveys of the
best SIM targets. For these surveys, SIM detects nearly all planets
above 9, 5, and 3 $M_{\earth}$, respectively. Figures~\ref{fig:213}B
and ~\ref{fig:214}B show completeness for Deep and Ultra-Deep
surveys of the best TPF-C targets. For these surveys, SIM finds
nearly all the terrestrial planets more massive than 7 and 4
$M_{\earth}$, respectively.

One important conclusion is that for all three survey modes, SIM
will find essentially all planets above 9 $M_{\earth}$ residing in
the habitable zones of target stars. This may be a significant
population of planets; simulations in Ida \& Lin support a maximum
terrestrial planet mass of up to $20 M_{\earth}$ for core accretion
models. Recently discovered Neptune-mass planets \citep{B2004,
M2004} may be the first indication of that population, since they
both reside inside the orbits of gas giants \citep{B2005B}.

\begin{figure}[p!]
\plotone{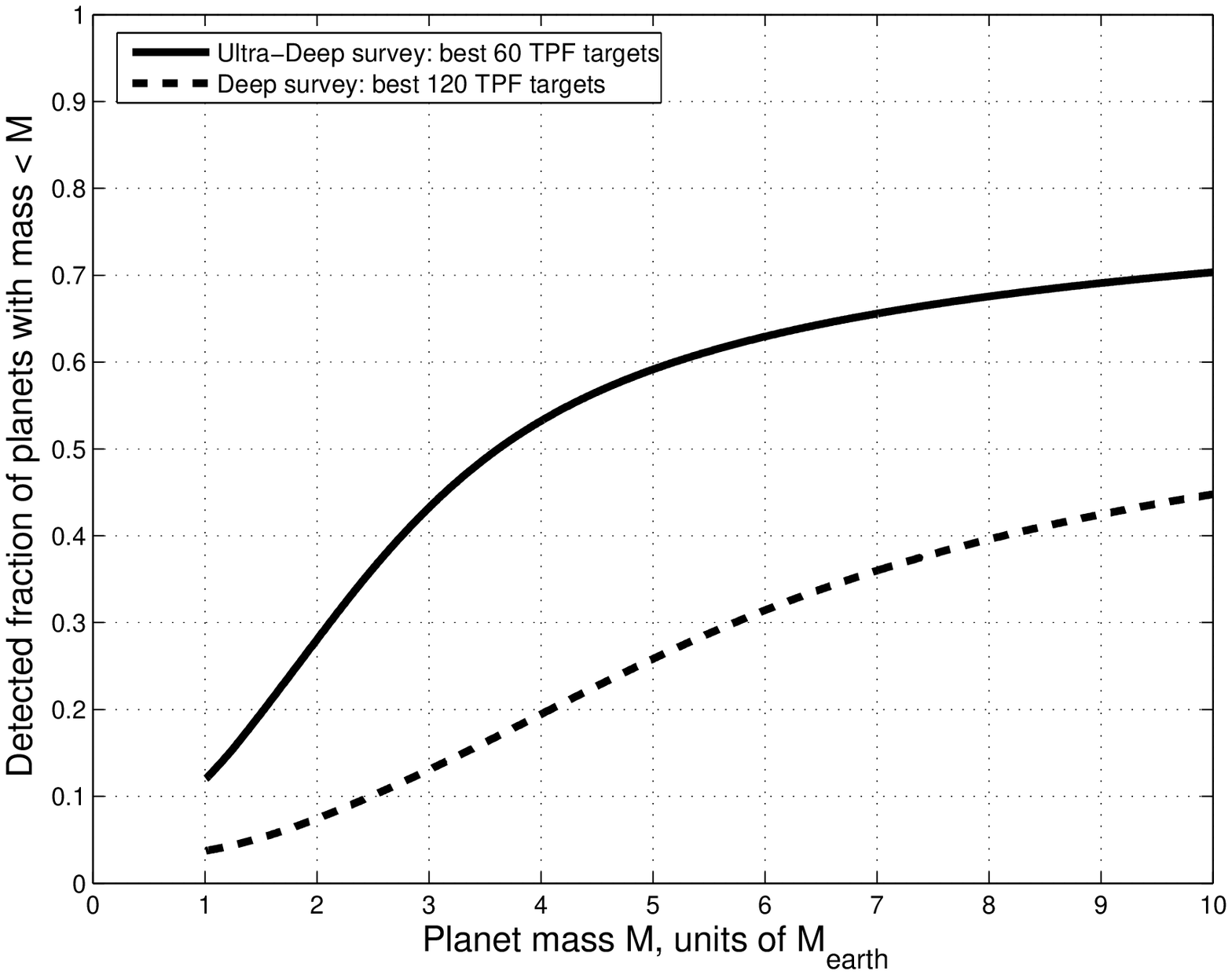} \caption{Cumulative completeness of SIM planet
discoveries, for surveys of the best TPF targets. Cumulative
completeness is the ratio of number of detected planets with mass $<
M$ to number of expected planets with mass $< M$. Single measurement
precision is $1.0~\mu as$, and detection threshold corresponds to
1\% false-alarm probability. Assumes that every target star has one
terrestrial planet and that the planet masses are distributed as
$M^{-1.1}$.} \label{fig:26}
\end{figure}

\begin{figure}[p!]
\plotone{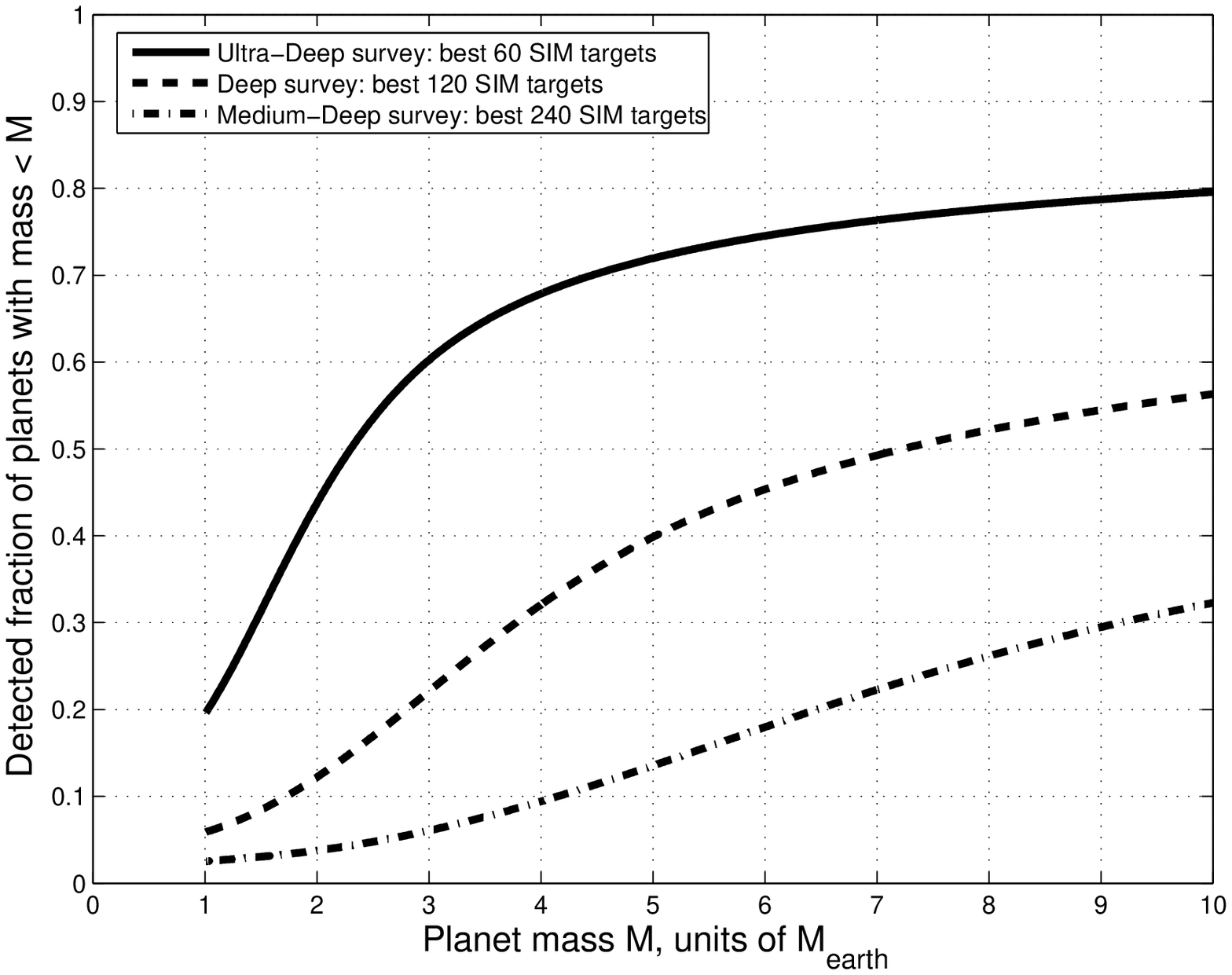} \caption{Cumulative completeness of SIM planet
discoveries, for surveys of the best SIM targets. Cumulative
completeness is the ratio of number of detected planets with mass $<
M$ to number of expected planets with mass $< M$. Single measurement
precision is $1.0~\mu as$, and detection threshold corresponds to
1\% false-alarm probability. Assumes that every target star has one
terrestrial planet and that the planet masses are distributed as
$M^{-1.1}$.} \label{fig:126}
\end{figure}

Cumulative completeness for surveys of the best TPF-C and SIM
targets is shown in Figures~\ref{fig:26} and ~\ref{fig:126},
respectively. We find that SIM will detect 32\%, 56\%, and 80\% of
terrestrial planets below $10 M_{\earth}$; and 6\%, 22\%, and 60\%
of terrestrial planets below $3 M_{\earth}$ for the Medium-Deep,
Deep and Ultra-Deep surveys, respectively.

\section{How SIM discoveries will benefit TPF}
\label{sec:Enrichment} The Terrestrial Planet Finder (TPF) mission
is being designed to have the capability to directly detect
terrestrial planets in the habitable zones of nearby stars. In
particular, the coronograph mission TPF-C shall be able to detect a
potentially habitable planet around at least 100 nearby stars
\citep{TPF2005}. While TPF-C can do this in the absence of apriori
information, achievement of its scientific priorities will
undoubtedly be furthered by knowledge of planetary statistics from
Kepler, and detections of terrestrial planets by SIM.

SIM's observations of TPF-C targets provide potentially valuable
information for the TPF mission. SIM's minimum detectable mass for a
target provides a lower limit for a detected planet's mass, while
the orbit period provides the star-planet separation. SIM will also
yield information about the \emph{confidence in a detection} -- the
probability that a planet is present in the case of a positive
detection; and the \emph{confidence in a non-detection} -- the
probability that a planet is absent, in the case of a non-detection.
It is straightforward to apply Bayes' rule to determine these
quantities. Results (averaged over the best 120 TPF-C targets
observed in a Deep survey) are shown in Figure~\ref{fig:17}. See
Appendix \ref{sec:appendix} for a discussion.

If SIM detects a planet at a given candidate TPF-C target star, the
minimum planet mass detectable by SIM, the star-planet separation,
together with the degree of confidence that a terrestrial planet is
present allows the TPF mission to assign a quantitative priority for
observing this potential target. On the other hand, if no planet is
detected by SIM, then the degree of confidence that no terrestrial
planet exists, together with SIM's and TPF's minimum detectable
masses again can help prioritize this target for TPF. For example,
at a given target, if the confidence that no terrestrial planet
exists is high, and TPF's minimum detectable mass is higher than
SIM's, the target would be given a very low priority.

For many stars with terrestrial planets, SIM will also characterize
the orbits, providing estimates of inclination, eccentricity, and
semimajor axis, allowing specification of optimal times and mirror
orientations for TPF observations. Because SIM detects planets
dynamically, it can unambiguously measure the planet's mass. SIM's
potential for astrometric orbit characterization and planet mass
determination has been addressed comprehensively in \citep{FT2003,
S2003, S2002}.

\section{Summary and Conclusions}
\label{sec:Summary}

We introduced the joint periodogram and showed that it is more
sensitive than the $\chi^2$ test for the null hypothesis for planet
detection. We derived a semi-empirical relation for minimum
detectable planet mass in terms of stellar distance and luminosity,
number of measurements over five years, and instrument noise. We
showed how this relation can be corrected for the effects of
parallax and proper motion. Using actual SIM target lists we
determined SIM's sensitivity for detection of terrestrial planets in
the habitable zones of nearby stars.  For the Medium-Deep, Deep, and
Ultra-deep surveys of the best SIM target stars within 30 pc,
we\begin{enumerate}
\item Evaluated the median minimum detectable mass for a planet at
mid-habitable zone. For surveys of the best 240, 120, and 60 SIM
target stars, we determine median minimum detectable planet masses
of 5.3, 2.9 and 1.6 $M_{\earth}$, respectively.
\item Determined the expected mass distribution and total number of
terrestrial planets that SIM will discover.  If each target star has
a terrestrial planet orbiting within its habitable zone, we find
that for surveys of the best 240, 120, and 60 SIM target stars,  SIM
will discover 77, 68, and 48 terrestrial planets, with mean mass of
6.2, 5.2, and 4.3 $M_{\earth}$ respectively; of these, 7, 13, and 18
planets, respectively, will have mass below $3 M_{\earth}$.
\item Determined the completeness of SIM's terrestrial planet
discoveries (i.e., the ratio of detected planets to expected
planets) as a function of planet mass.  For the three specified
surveys of the best SIM targets, we find that SIM detects 32\%,
56\%, and 80\% of terrestrial planets below $10 M_{\earth}$; and
6\%, 22\%, and 60\% of terrestrial planets below $3 M_{\earth}$.
\end{enumerate}

Finally, we discussed the confidence in SIM detections and
non-detections, and described how information from SIM's planet
survey can enable Terrestrial Planet Finder (TPF) to increase its
yield of terrestrial planets.

\noindent We are grateful to Serge Dubovitsky, Debra Fischer, Chris
Gelino, Andy Gould, Geoff Marcy, Chris McCarthy, Bijan Nemati, and
Stuart Shaklan for helpful discussions and/or other contributions to
this work. We wish to thank an anonymous referee for a most thorough
critical review, with many recommendations that greatly improved the
paper. This work was carried out at the Jet Propulsion Laboratory,
California Institute of Technology, under contract with NASA.

\appendix
\section{Confidence in a detection or non-detection}\label{sec:appendix}
Given the occurrence rate $\eta_{terrestrial}$ for terrestrial
planets, the false-alarm probability  $F$ at the detection
threshold, and the probability $P_{det}$ that an existing
terrestrial planet will be detected, we can use Bayes' Rule to
determine the confidence in a detection (i.e., the probability that
a detection is not a false positive) at a given target star.

Formally, Bayes' rule for the confidence in a detection is:
\begin{equation}\label{eq:A1}p(TP~exists | TP~detected) = \frac{p(TP~
detected | TP~exists) \times p(TP~exists)}{p(TP~
detected)},\end{equation} where TP stands for terrestrial planet and
(in accordance with standard usage) the symbol $|$ means
\emph{conditioned on}, or \emph{given}. The denominator of the
right-hand side of equation (\ref{eq:A1}) can be expanded as
$p(TP~detected) = p(TP~detected|TP~exists)\times p(TP~exists) +
p(TP~detected|TP~\overline{exist})\times p(TP~\overline{exist})$.
Recognizing that $p(TP~detected|TP~\overline{exist}) \equiv F$,
 $p(TP~detected|TP~exists)\equiv P_{det}$, $p(TP~exists) \equiv
\eta_{terrestrial}$ and $p(TP~\overline{exist})\equiv 1 -
\eta_{terrestrial}$, we rewrite the denominator of the right-hand
side of Equation (\ref{eq:A1}) as $p(TP~detected) =
P_{det}\times\eta_{terrestrial} + F\times(1-\eta_{terrestrial})$,
and the numerator as $P_{det}\times \eta_{terrestrial}$.  The
confidence in a detection is therefore
\begin{equation}\label{eq:A3}p(TP~exists | TP~detected) = \frac{P_{det}\times
\eta_{terrestrial}}{P_{det}\times\eta_{terrestrial} +
F\times(1-\eta_{terrestrial})}\end{equation}

Similarly, the confidence in a non-detection (i.e., the probability
that a non-detection is not a missed detection) at a given target
star can also be expressed in terms of Bayes' rule, as follows:
\begin{equation}\label{eq:A2}p(TP~\overline{exist} | TP~\overline{detected}) = \frac{p(TP~
\overline{detected} | TP~\overline{exist}) \times
p(TP~\overline{exist})}{p(TP~ \overline{detected})}\end{equation}
The denominator of the right-hand side of (\ref{eq:A2}) can be
expanded as $p(TP~\overline{detected}) =
p(TP~\overline{detected}|TP~\overline{exist})\times
p(TP~\overline{exist}) + p(TP~\overline{detected}|TP~exists)\times
p(TP~exists)$. Finally, since
$p(TP~\overline{detected}|TP~\overline{exist})\equiv 1-F$, and
$p(TP~\overline{detected}|TP~exists)\equiv 1-P_{det}$, the
confidence in a non-detection becomes
\begin{equation}\label{eq:A4}p(TP~\overline{exist} |
TP~\overline{detected}) =
\frac{(1-F)\times(1-\eta_{terrestrial})}{(1-F)\times(1-\eta_{terrestrial})+(1-P_{det})\times\eta_{terrestrial}}\end{equation}

\begin{figure}[p!]
\plottwo{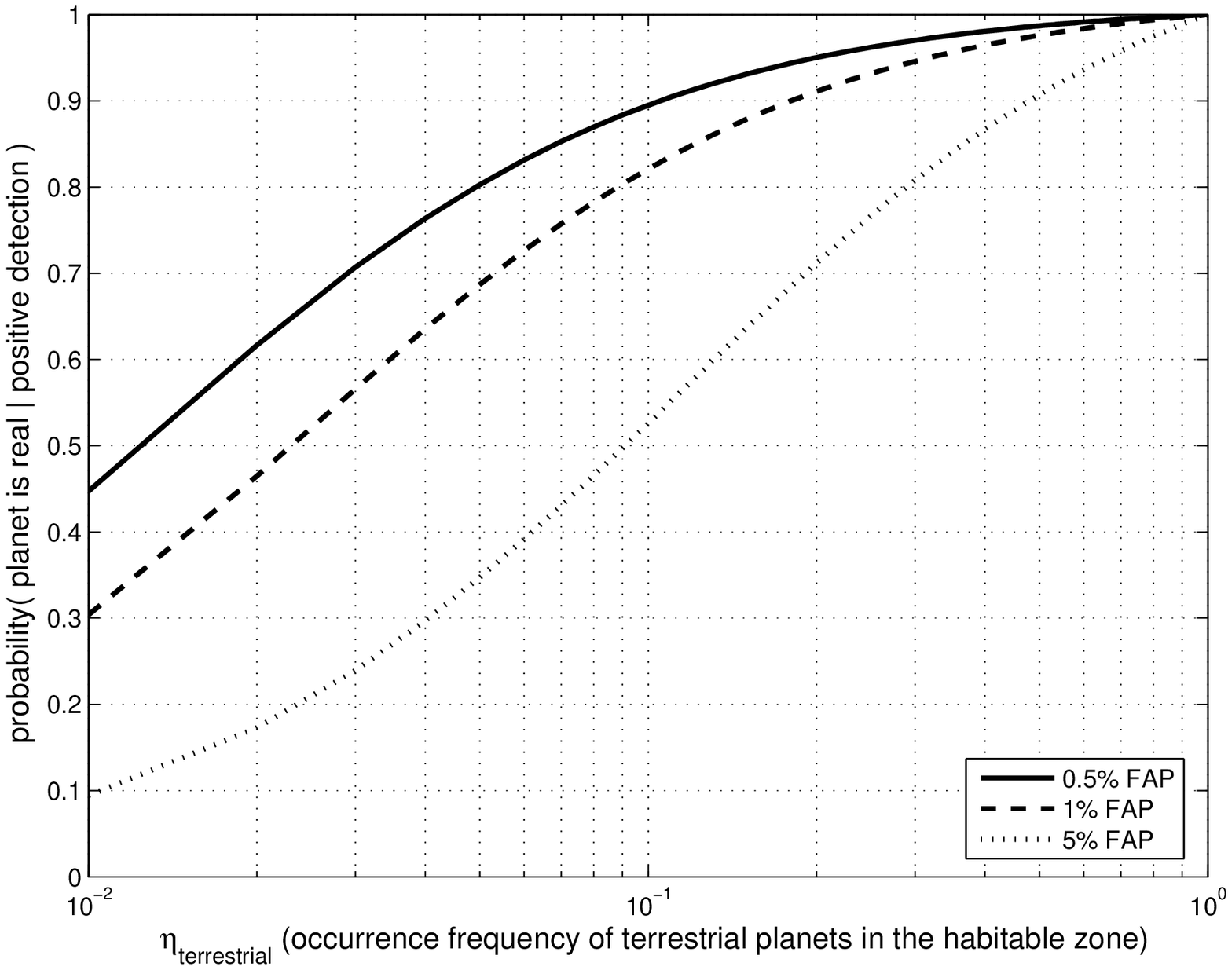}{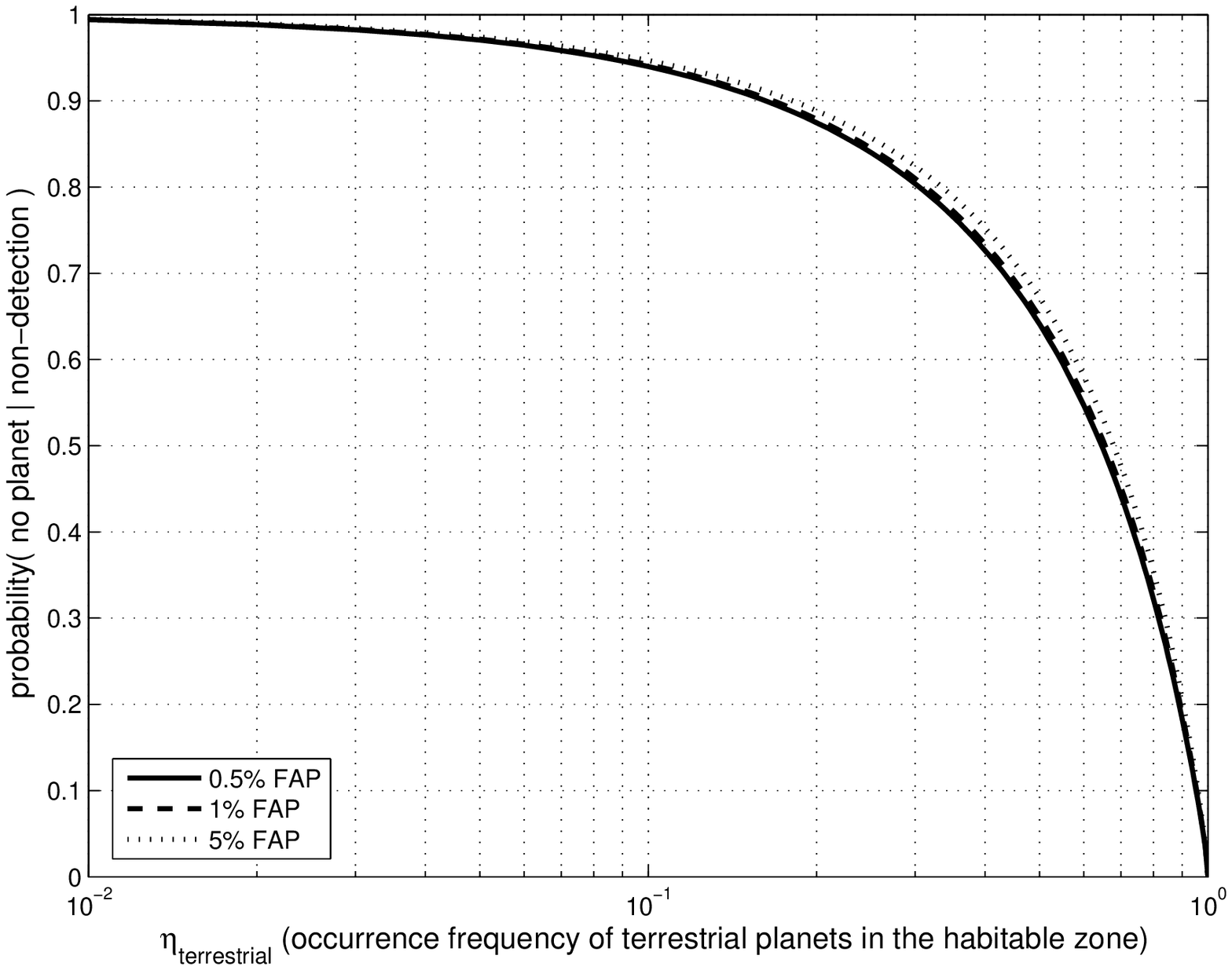}\caption{ Left: Confidence in
the \emph{presence} of a terrestrial planet, given a detection.
Right: Confidence in the \emph{absence} of a terrestrial planet,
given a non-detection. Both plots are for Deep planet survey of best
120 TPF targets -- 104 two-dimensional measurements per star, at
single measurement precision of $1.0~\mu as$, and show results for
detection thresholds corresponding to false-alarm probabilities of
0.5\%, 1\%, and 5\%. Planet mass distribution is assumed
$\varpropto~M^{-1.1}$. Abscissa is the occurrence rate of
terrestrial planets in the habitable zone. Results shown are
averaged over all the stars in the survey.} \label{fig:17}
\end{figure}

In practice the confidences would be computed on a star-by-star
basis. Figure~\ref{fig:17} shows confidences in a detection and in a
non-detection, respectively, averaged over all targets, for a Deep
survey of the best 120 TPF-C targets. If $\eta_{terrestrial}$ is
0.1, then for detection threshold corresponding to 1\% false-alarm
probability, average confidence in a detection is 82\% and average
confidence in a non-detection is 94\%. The former result means that
we are 82\% certain, on average, that a detection is really a
terrestrial planet in the habitable zone and not a false positive.
The latter result means that on average, we are 94\% certain that a
non-detection rules out the existence of terrestrial planet in the
habitable zone.

\begin{deluxetable}{llllll}
\tablecaption{Number of Detected Planets versus Minimum Mass
\tablenotemark{\dagger} } \tablewidth{0pt} \tablehead{ \colhead{Min
Mass} & \colhead{SIM Ultra-Deep} & \colhead{SIM Deep} & \colhead{SIM
Medium-Deep} & \colhead{TPF Ultra-Deep} & \colhead{TPF Deep}}
\startdata
10 $M_{\earth}$ &  60 & 120 & 240   &  60   & 120\\
9               &  60 & 120 & 240   &  60   & 120\\
8               &  60 & 120 & 240   &  60   & 120\\
7               &  60 & 120 & 240   &  60   & 120\\
6               &  60 & 120 & 179   &  60   & 120\\
5               &  60 & 119 & 101   &  60   & 110\\
4               &  60 & 114 & 55    &  60   & 62\\
3               &  60 & 64  & 23    &  60   & 34\\
2               &  55 & 20  & 6     &  30   & 11\\
1.5             &  23 & 7   & 4     &  12   & 4\\
1               &  6  & 4   & 3     &  3    & 2\\
\label{tbl:1}
\enddata
\tablenotetext{\dagger}{For 50\% detection efficiency and 1\%
false-alarm probability}
\end{deluxetable}

\begin{deluxetable}{lcccc}
\tablecaption{SIM terrestrial planet
discoveries\tablenotemark{\dagger}} \tablewidth{0pt} \tablehead{
\colhead{Survey} & \colhead{Median min. detectable mass}&
\colhead{$< 10M_{\earth}$} & \colhead{$< 3M_{\earth}$ } &
\colhead{Mean mass} } \startdata
Best 60 for SIM  & 1.6 $M_{\earth}$ & 48 & 18 &  4.3 $M_{\earth}$\\
Best 60 for TPF  & 2.0 $M_{\earth}$& 42 & 13 &  4.6 $M_{\earth}$\\
Best 120 for SIM & 2.9 $M_{\earth}$& 68 & 13 &  5.2 $M_{\earth}$\\
Best 120 for TPF & 4.0 $M_{\earth}$& 54 & 8  &  5.6 $M_{\earth}$\\
Best 240 for SIM & 5.3 $M_{\earth}$& 77 & 7  &  6.2 $M_{\earth}$\\
\label{tbl:2}
\enddata
\tablenotetext{\dagger}{Assuming each target has a terrestrial
planet, with masses distrubuted as $M^{-1.1}$ \cite{TT2002} }
\end{deluxetable}

\clearpage

\clearpage

\clearpage

\clearpage

\clearpage

\clearpage

\clearpage

\clearpage

\clearpage

\clearpage

\clearpage

\clearpage

\clearpage

\clearpage

\clearpage

\clearpage

\clearpage

\clearpage

\clearpage

\clearpage

\clearpage

\end{document}